\documentclass[journal]{vgtc}              

\onlineid{1399}

\vgtccategory{Research}

\vgtcpapertype{application/design study}

\title{HypoChainer: A Collaborative System Combining LLMs and Knowledge Graphs for Hypothesis-Driven Scientific Discovery}

\author{
  \authororcid{Haoran Jiang}{0009-0009-5717-4208},
  \authororcid{Shaohan Shi}{0009-0004-3384-8304},
  \authororcid{Yunjie Yao}{0009-0000-8371-6984},  
  \authororcid{Chang Jiang}{0000-0002-7468-3372}, 
  \authororcid{Quan Li}{0000-0003-2249-0728}
}

\authorfooter{
H. Jiang, S. Shi, Y. Jie and Q. Li (corresponding author) are with School of Information Science and Technology, ShanghaiTech University, and Shanghai Engineering Research Center of Intelligent Vision and Imaging, China. E-mail: \{jianghr2023, shishh2023, yaoyj2024, liquan\}@shanghaitech.edu.cn. 
C. Jiang is with the Shanghai Clinical Research and Trial Center, Shanghai, China. E-mail: jiangchang@shanghaitech.edu.cn.
Haoran Jiang and Shaohan Shi contributed equally to this work.}
\abstract{Modern scientific discovery faces challenges in integrating the rapidly expanding and diverse knowledge required for exploring novel knowledge in biology. While traditional hypothesis-driven research has proven effective, it is constrained by human cognitive limitations, knowledge complexity, and the high costs of trial-and-error experimentation. Deep learning models, particularly graph neural networks (GNNs), have accelerated scientific progress. However, the vast predictions generated make manual selection for experimental validation impractical. Attempts to leverage large language models (LLMs) for filtering predictions and generating novel hypotheses have been impeded by issues such as hallucinations and the lack of structured knowledge grounding, which undermine their reliability. To address these challenges, we propose \textit{HypoChainer}, a collaborative visualization framework that integrates human expertise, LLM-driven reasoning, and knowledge graphs (KGs) to enhance scientific discovery visually. \textit{HypoChainer} operates through three key stages: (1) \textit{Contextual Exploration}: Domain experts employ retrieval-augmented LLMs (RAGs) and visualizations to extract insights and research focuses from vast GNN predictions, supplemented by interactive explanations for in-depth understanding; (2) \textit{Hypothesis Construction}: Experts iteratively explore the KG information relevant to the predictions and hypothesis-aligned entities, gaining knowledge and insights while refining the hypothesis through suggestions from LLMs; and (3) \textit{Validation Selection}: Predictions are prioritized based on the refined hypothesis chains and KG-supported evidence, identifying high-priority candidates for validation. The hypothesis chains are further optimized through visual analytics of the retrieval results. We evaluated the effectiveness of \textit{HypoChainer} in hypothesis construction and scientific discovery through a case study and expert interviews.}


\teaser{
  \centering
  \includegraphics[width=\linewidth]{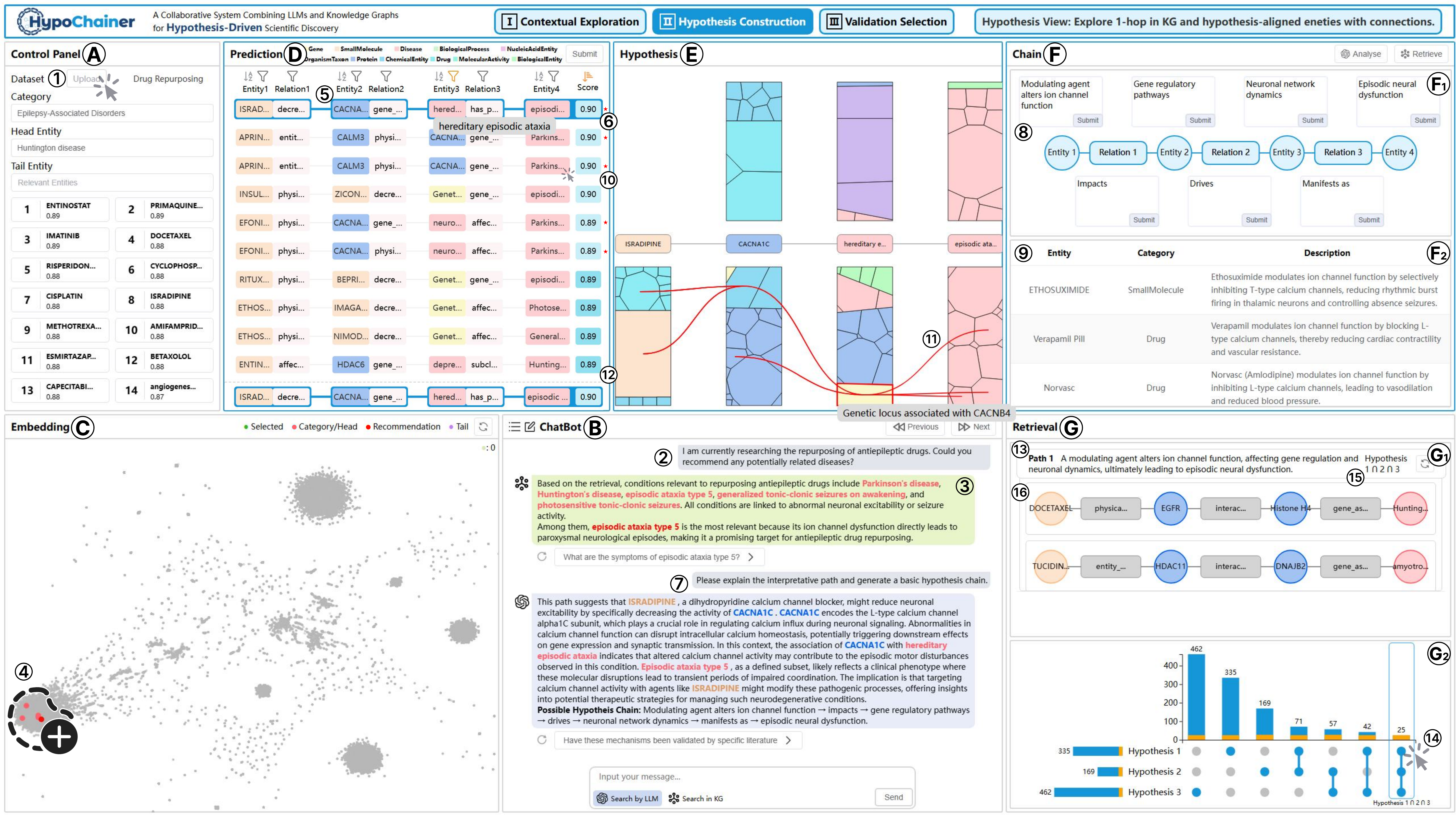}
  \vspace{-6mm}
  \caption{In the presented case study, the biologist's analytical workflow unfolds as follows: \protect\circled{1} Upload Drug Repurposing data. \protect\circled{2} Pose a target question to the RAG model. \protect\circled{3} RAG identifies \textit{Episodic Ataxia Type 5} as the top candidate. \protect\circled{4} Lasso the clustered diseases. \protect\circled{5} Observe that the \textit{CACNA1C} often appeared among the top-ranked predictions. \protect\circled{6} Select the interpretative path of \textit{Episodic Ataxia Type 5}. \protect\circled{7} The LLM explains the selected prediction path and generates base hypotheses. \protect\circled{8} Construct the hypothesis chain. \protect\circled{9} Hypothesis chain is validated for entity alignment. \protect\circled{10} KG integration reveals novel paths in the lower layer. \protect\circled{11} Observe multiple interpretative paths consistent with the hypothesis. \protect\circled{12} Inconsistent predictions for \textit{Huntington's disease} trigger afterwards exploration. \protect\circled{13} Expand the retrieval results. \protect\circled{14} - \protect\circled{15} Filtered predictions confirm relevance to \textit{Parkinson's disease} and \textit{Amyotrophic Lateral Sclerosis} (ALS). \protect\circled{16} Identify overlooked predictions of \textit{Huntington's disease} aligning with the refined hypothesis chain.}
  \label{fig:teaser}
}

\graphicspath{{figs/}{figures/}{pictures/}{images/}{./}} 

\usepackage{tabu}                      
\usepackage{booktabs}                  
\usepackage{lipsum}                    
\usepackage{mwe}                       
\usepackage{tabu}                      
\usepackage{booktabs}                  
\usepackage{lipsum}                    
\usepackage{mwe}                       
\usepackage{multirow}
\usepackage{wrapfig}
\usepackage[dvipsnames,svgnames]{xcolor}
\usepackage{times}                     
\usepackage{cite}       
\usepackage{paralist}
\usepackage{balance}

\usepackage{amsthm,amsmath,amssymb}
\usepackage{tikz}

\usepackage{mathrsfs}
\usepackage{pifont}

\usepackage{mathptmx}                  
\usepackage[many]{tcolorbox}

\newcommand{\review}{\textcolor{black}}

\definecolor{mycustomcolor}{HTML}{2D2C4C}
\definecolor{casecolor}{HTML}{E5E5E5}
\definecolor{graycolor}{HTML}{888888}

\definecolor{correct}{HTML}{A52A2A}

\definecolor{RWcolor}{RGB}{26, 67, 120}
\definecolor{RWcolor2}{RGB}{20, 150, 217}

\newtcbox{\RWbox}[1][RWcolor]
  {on line, arc = 2pt, outer arc = 3pt,
    colback = RWcolor2!10!white, colframe = RWcolor,
    boxsep = 0pt, left = 1pt, right = 1pt, top = 1pt, bottom = 1pt,
    boxrule = 1pt, bottomrule = 1pt, toprule = 1pt}

\newcommand*\caserectangle[1]{\tikz[baseline=(char.base)]{
            \node[rectangle, draw=black, fill=white, text=black, inner sep=2pt] (char) {#1};}}

\newcommand*\alterrectanglered[1]{\tikz[baseline=(char.base)]{
            \node[rectangle, draw=red, fill=white, text=red, inner sep=2pt, rounded corners=2pt] (char) {#1};}}

\newcommand*\alterrectangleblue[1]{\tikz[baseline=(char.base)]{
            \node[rectangle, draw=blue, fill=white, text=blue, inner sep=2pt, rounded corners=2pt] (char) {#1};}}

\newcommand*\whiterectangle[1]{\tikz[baseline=(char.base)]{
            \node[rectangle, draw=black, fill=white, text=black, inner sep=2pt] (char) {#1};}}

\newcommand*\grayrectangle[1]{\tikz[baseline=(char.base)]{
            \node[rectangle, draw=black, fill=white, text=black, inner sep=2pt] (char) {#1};}}

\usepackage{xstring}
\usepackage{tikz}
\newcommand*\midfontsize{\fontsize{7pt}{7pt}\selectfont}

\newcommand*\circled[1]{%
  \StrLen{#1}[\mylength]%
  \ifnum\mylength=1
    \tikz[baseline=(char.base)]{
      \node[shape=circle,draw=black,fill=white,inner sep=0.5pt,
            line width=0.5pt,text=black,font=\footnotesize] (char) {#1};%
    }%
  \else
    \tikz[baseline=(char.base)]{
      \node[shape=circle,draw=black,fill=white,inner sep=0.1pt,
            line width=0.1pt,text=black,font=\midfontsize] (char) {#1};%
    }%
  \fi
}

\usepackage{titlesec}

\newcommand*\specialcircledg[2]{%
\tikz[baseline=(char.base)]{
    \node[shape=circle,draw=black,fill=white,text=black,align=center,inner sep=0.5pt, font=\footnotesize, line width=0.5pt] (char) {%
        \scalebox{0.8}{#1}\scalebox{0.6}{#2}};}%
}

\newcommand*\specialcircledgsingle[1]{%
\tikz[baseline=(char.base)]{
    \node[shape=circle,draw=black,fill=white,align=center,text=black,inner sep=0.5pt, font=\footnotesize, line width=0.5pt] (char) {#1};}%
    }

\definecolor{mybgcolorHHH}{HTML}{CAD8FF} 
\definecolor{mybgcolorLHH}{HTML}{CFDFFF} 
\definecolor{mybgcolorLLH}{HTML}{E9F1FF} 
\definecolor{mybgcolorLLL}{HTML}{F0F8FF} 

\definecolor{gene}{HTML}{9ECA80}
\definecolor{cc}{HTML}{FFC488}
\definecolor{bp}{HTML}{98CAF7}
\definecolor{pathway}{HTML}{CDC0DB}
\definecolor{mf}{HTML}{F7A8C3}

\begin{document}
\maketitle

\section{Introduction}

\par
\begin{figure}[h]
\centering
\vspace{-3mm}
\includegraphics[width=\linewidth]{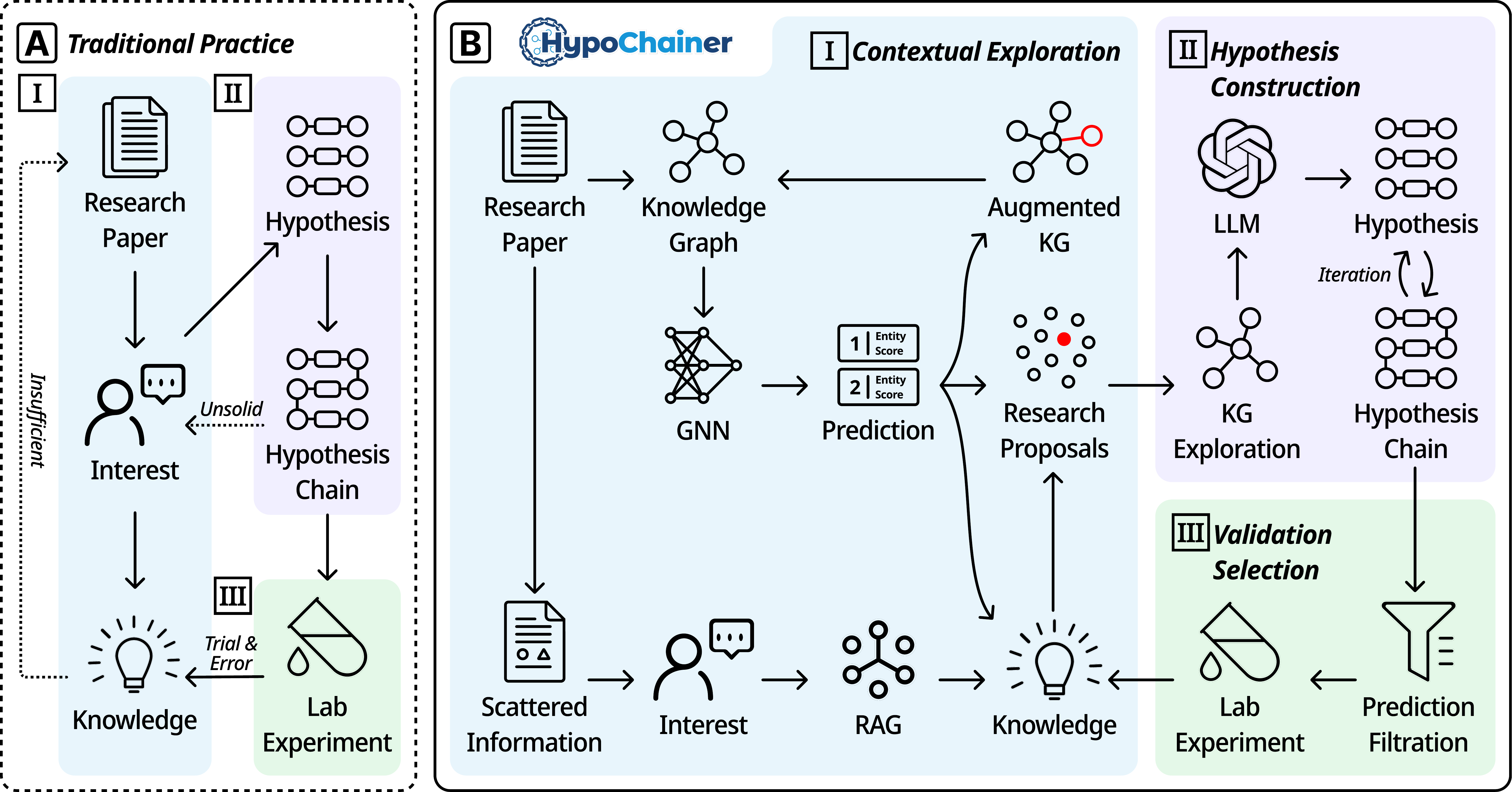}
\vspace{-6mm}
\caption{Comparison of Traditional Practice \protect\grayrectangle{A} and \textit{HypoChainer} Pipeline \protect\grayrectangle{B}: Both follow the \protect\whiterectangle{I} \textbf{Contextual Exploration}, \protect\whiterectangle{II} \textbf{Hypothesis Construction}, and \protect\whiterectangle{III} \textbf{Validation Selection} workflow.}
\label{fig:pipeline}
\vspace{-3mm}
\end{figure}

 Modern scientific discovery faces a critical challenge in synthesizing exponentially growing~\cite{5958175}, heterogeneous knowledge to drive breakthroughs in data-intensive domains~\cite{ghafarollahi2024sciagents}, like biomedicine and drug development~\cite{zhou2024hypothesis}. Traditional hypothesis-driven research (\cref{fig:pipeline}-\grayrectangle{A}) relies on domain experts to manually formulate theories through exhaustive literature reviews, iterative experimentation, and reasoning grounded in specialized knowledge~\cite{popper2005logic}. While this paradigm has yielded significant advances, it's inherently constrained by cognitive limits~\cite{9623273}, combinatorial complexity of biological systems~\cite{birtwistle2015analytical}, and resource-intensive nature of experimental validation (e.g., months-long wet-lab studies for a single hypothesis)~\cite{yang2024moose}. These limitations highlight the need for computational frameworks that enhance human expertise by reducing information overload and reliance on trial-and-error.

\par Advances in deep learning, particularly graph neural networks (GNNs) and transformer-based models~\cite{zhang2023kr4sl,yao2024km}, have revolutionized \review{hypothesis prioritization—the process of ranking and selecting the most promising hypotheses—in biological discovery.} These methods effectively model biological interactions---such as protein-ligand binding~\cite{dang2017biolinker}, synthetic lethality (SL) relationships\footnote{\small{Synthetic lethality is a genetic interaction where the simultaneous inhibition of two genes causes specific cell death while inhibiting either gene alone does not.}}~\cite{zhang2023kr4sl}, and drug candidate prediction~\cite{ma2023kgml}---to generate predictive outcomes that guide hypothesis validation. While these tools have greatly streamlined the scientific discovery process, their effectiveness is increasingly constrained by the rapid expansion of biological and biomedical datasets. As model predictions grow in volume~\cite{ma2023kgml}, manual evaluation and validation by domain experts becomes impractical, creating a critical bottleneck in translating computational insights into scientific breakthroughs. To address this challenge, previous studies have proposed using the comparative analysis of similar predictive outcomes~\cite{jiang2024slinterpreter} to identify promising yet unverified predictions. Although these methods have shown success in identifying results that align with established mechanisms, they are limited in their ability to detect predictions that lack prior validated analogs. As a result, the identification and validation of novel mechanisms from predictive outcomes remain a significant challenge, hindering the discovery of groundbreaking scientific insights.

\par Recent efforts~\cite{yang2024unleashing,ahmed2025leveraging, zheng2025automationautonomysurveylarge} have extended to leveraging large language models (LLMs) to analyze and interpret large-scale predictive outcomes. LLMs offer unique advantages in integrating multimodal and heterogeneous data, enabling preliminary reasoning frameworks to address complex biological questions~\cite{naveed2024comprehensiveoverviewlargelanguage,app14125068}---from hypothesis generation and information retrieval~\cite{10.1007/978-981-97-9434-8_33} to evidence-based explanations~\cite{schimanski-etal-2024-towards}. Yet, as data complexity grows, their propensity for hallucinations or inaccuracies intensifies~\cite{10569238}, raising concerns about reliability despite their multi-perspective reasoning capabilities. This tension has spurred interest in grounding LLM outputs in structured knowledge~\cite{10.5555/3491440.3491820,10.1145/3340531.3411975} to enhance trustworthiness, where GNNs and knowledge graphs (KGs) have emerged as critical solutions. Particularly, GNNs, renowned for accuracy through structured relational modeling, synergize with KGs---which encode precise, standardized relationships---to improve feature representation and prediction robustness~\cite{zhang2024multidomainknowledgegraphcollaborative}. For instance, \textit{KR4SL}~\cite{zhang2023kr4sl} integrates KG reasoning with GNNs to predict synthetic lethality, combining semantic relationships and structural patterns to boost both accuracy and explainability. However, challenges persist: textual data alone may inadequately contextualize predictions, while gaps in commonsense knowledge or overly simplistic edge relationships in KGs can limit their comprehensiveness~\cite{peng2023knowledge}.

\par To leverage the complementary strengths and mitigate the limitations of existing methods (LLMs, KGs, and manual workflows), we focus on a critical challenge: \textit{integrating human expertise, LLM-driven reasoning, and KG-structured knowledge} into a unified hypothesis-driven framework that breaks information silos and streamlines the discovery of novel mechanisms. 
Through a formative study conducted in collaboration with domain experts, we identified six key design requirements, emphasizing collaborative exploration, interpretability, and iterative hypothesis refinement. Guided by these requirements, we propose a collaborative framework that synergizes human intuition, LLMs, and KGs through \review{the construction of hypothesis chains—structured reasoning paths composed of multiple interrelated hypotheses connected by logical links.} The workflow unfolds in three stages: 
\whiterectangle{I} \textbf{Contextual Exploration}: Domain experts raise research questions, prompting a retrieval-augmented LLM (RAG) to surface relevant research objects from GNN predictions. Interactive visualizations and LLM-generated explanations contextualize results, enabling experts to better analyze predictions while addressing gaps in the missing commonsense knowledge and details within structured information. \whiterectangle{II} \textbf{Hypothesis Construction}: As experts iteratively analyze predictions, they construct hypothesis chains---semantically linked sequences of insights---supported by LLM-generated refinements and KG relationships. \whiterectangle{III} \textbf{Validation Selection}: The workflow filters predictions against refined hypothesis chains, identifying candidates for experimental validation based on alignment with KG-supported evidence. Weak points in the hypothesis chain are further optimized through visual analytics of the retrieval results.
To demonstrate the effectiveness of \textit{HypoChainer} (\cref{fig:pipeline}-\grayrectangle{B}), we conducted expert interviews and a case study in the field of drug repurposing. In summary, the contributions of this study are as follows:
\begin{compactitem}
    \item We conducted in-depth expert interviews and a thorough literature analysis, identifying six key design requirements for integrating LLMs with structured knowledge in \review{scientific discovery}.
    \item We developed a collaborative framework, \textit{HypoChainer}, linking LLMs and KGs, enabling experts to explore model predictions, construct and refine hypothesis chains, uncover underlying mechanisms, and prioritize validations through interactive visualization.
    \item We validated \textit{HypoChainer} through a comprehensive case study and expert interviews, demonstrating its effectiveness and generalizability in hypothesis construction and scientific discovery.

\end{compactitem}

\section{Related work}
\subsection{Hypothesis Generation and Refinement}
\par Hypothesis generation involves proposing new concepts or scientific mechanisms~\cite{zhou2025hypothesis}, playing a vital role in advancing scientific research~\cite{qi2023large}, which relies heavily on researchers' accumulated knowledge and intuition, introducing limitations and uncertainties. To address these challenges, researchers began leveraging extensive existing knowledge to support hypothesis construction. Early hypothesis generation efforts primarily focused on predicting relationships between concepts, based on the assumption that new ideas emerged from connections with established ones~\cite{henry2017literature,krenn2022predicting}. However, with advancements in language models~\cite{zhao2023survey,chang2024survey}, open-ended idea generation has gained increasing attention~\cite{si2024can,kumar2024can}. Recent AI-driven hypothesis generation methods employ diverse approaches to conceptualizing research ideas. For instance, \textit{MOOSE}~\cite{yang2024moose} and \textit{IdeaSynth}~\cite{pu2024ideasynth} integrate LLMs into interactive frameworks, facilitating the transition from inspiration to hypothesis construction. Many studies~\cite{Kehrer2008Robustness,Vohra2023Neural,shi2022medchemlens} have also combined hypothesis-driven frameworks with visualization-based designs. 

\par Moreover, hypothesis generation is rarely a one-step process, particularly when constructing complex hypothesis chains that require logical consistency and adherence to fundamental principles. Therefore, hypothesis refinement---driven by feedback and iterative improvements---is equally critical. Methods such as \textit{HypoGeniC}~\cite{zhou2024hypothesis} and \textit{MOOSE}~\cite{yang2024moose} emphasize iterative enhancement through feedback mechanisms, including direct responses to hypothesis~\cite{baek2024researchagent}, experimental result evaluations~\cite{ma2024llm,yuan2025dolphin}, and automated peer-review commentary~\cite{lu2024ai}. Beyond feedback-driven refinements, collaborative hypothesis generation has also gained traction, leading to the development of multi-agent systems~\cite{nigam2024acceleron,ghafarollahi2024sciagents}. For instance, \textit{VIRSCI}~\cite{su2024two} optimizes hypothesis construction by customizing knowledge for each agent, while \textit{Nova}~\cite{hu2024nova} incorporates outputs from other research efforts to refine hypothesis generation. However, such multi-agent frameworks also introduce challenges, including hallucinations~\cite{10569238}, inaccuracies, and errors in agent-generated outputs. If left unchecked, these errors can propagate through the reasoning process, leading to misleading or incorrect conclusions. Recently, a multi-agent system based on \textit{Gemini 2.0} was proposed~\cite{gottweis2025towards}, designed to generate and refine novel research hypotheses through a ``generate-debate-evolve'' framework, utilizing test-time compute to improve hypothesis quality. However, this system does not incorporate constraints from structured knowledge sources and experts still remain limited to guiding AI in knowledge discovery, lacking intuitive visualizations to independently explore and uncover insights.

\par To address these challenges, we propose combining LLM-driven hypothesis generation with AI-expert collaborative optimization process. By enabling timely human intervention and error correction, it ensures both the logical consistency of the hypothesis and the effective utilization of LLM's vast knowledge and reasoning capabilities, ultimately enhancing hypothesis construction and iterative refinement.

\subsection{Graph-Structured Knowledge Exploration}
\par With the continuous expansion and enrichment of structured knowledge, increasingly complex graphs have emerged across various domains. These graphs encode rich information about entities and their relationships, facilitating reasoning and novel knowledge prediction through propagation. However, as graph data rapidly expands, effectively identifying patterns within large-scale graphs and exploring vast prediction results within the same large-scale KG has become increasingly challenging. Researchers have addressed these issues by developing various visualizations, particularly in biological networks~\cite{jiang2024slinterpreter,wang2022extending}, neural networks~\cite{jin2022gnnlens}, and social networks~\cite{Chen2020R-Map}. For example, Paley et al.~\cite{paley2021pathway} used force-directed layouts and clustering to reveal key biological pathways and drug targets in complex datasets, facilitating discoveries in molecular interactions and metabolic pathways that drive advancements in drug discovery and synthetic biology.

\par While these visualization methods have demonstrated significant effectiveness, the challenges of graph exploration and information retrieval extend beyond static representations. Many interactive approaches~\cite{zheng2024disciplink} have been developed to tackle these challenges, such as \textit{Biolinker}~\cite{dang2017biolinker}, an interactive visualization system that supports bottom-up exploration of complex protein interaction networks. However, when analyzing predictive model outputs, researchers must navigate large volumes of predictions and interpret extensive graph structures---often with subtle yet critical distinctions. These challenges are further exacerbated when investigating intricate and extended chains of interpretative paths and patterns, making graph analysis increasingly demanding.

\par Building upon previous research in graph-structured knowledge exploration and leveraging the analytical reasoning capabilities of LLMs, this study integrates graph exploration with text-based prompts. We employ a hierarchical layout to organize one-hop entities along prediction paths and \review{hypothesis-aligned entities—defined as those complying with or related to the given hypothesis descriptions—where hypotheses are initially generated by the LLM and iteratively refined by experts}, using Voronoi treemaps~\cite{balzer2005Voronoi}, enhancing spatial efficiency and clarity. Simultaneously, existing edges within the KG are visualized between entities to highlight structural relationships between predictions and entities from different sources. This dual approach enables users to explore relationships among structurally related and hypothesis-aligned entities, evaluate the coherence between model predictions and proposed hypotheses, and uncover insights through comparative analysis.

\subsection{Collaboration Between Human, LLM, and KG}
\par As AI advances and the modes of interaction diversify, the dynamics of both human-AI and AI-to-AI cooperation are continuously evolving. Within the framework of \review{\textit{Collaboration Between Human, LLMs, and KG}}, the pairwise interactions~\cite{kaufmann2023survey,amer2023large} can be categorized into three key relationships: \textbf{Human-LLM}, \textbf{Human-KG}, and \textbf{LLM-KG}.

\par \textbf{Human-LLM}. The interaction between humans and LLMs forms a continuous learning loop. In this collaborative paradigm, humans serve as both users and critical coordinators, guiding context-specific outputs through structured mechanisms. Additionally, reinforcement learning~\cite{6025669} from human feedback further refines LLM outputs to align with ethical considerations and domain-specific objectives~\cite{kaufmann2023survey}. 

\par \textbf{Human-KG}. Humans engage with KGs to retrieve, structure, and refine information through queries~\cite{perez2009semantics}, natural language~\cite{schneider2023data}, and visualization tools~\cite{asprino2021pattern}. By leveraging KGs as structured repositories of factual knowledge, users can enhance information accessibility and explainability in AI applications. For instance, \textit{DrugExplorer}~\cite{wang2022extending} employs explainable AI (XAI) techniques to interpret GNN-based predictions and refine biomedical KGs, while \textit{SLInterpreter}~\cite{jiang2024slinterpreter} introduces an iterative human-AI collaboration system for SL prediction, which enables experts to enhance KG interpretability and align AI outputs with expertise via metapath strategies and iterative path refinement. 

\par \textbf{LLM-KG}. The bidirectional integration between LLMs and KGs unfolds through two key paradigms: \textit{LLM-enhanced KG}~\cite{amer2023large} and \textit{KG-enhanced LLM}~\cite{pan2024unifying}. LLMs enhance KGs by automating \textit{entity extraction} via prompt-driven mining~\cite{jiang2020can}, refining \textit{entity parsing and matching} through synthetic labeled data generation~\cite{vrandevcic2014wikidata}, and improving \textit{link prediction} with joint text-KG embeddings~\cite{wang2021kepler}. Conversely, KGs enrich LLMs by injecting structured knowledge during \textit{pre-training}~\cite{he-etal-2021-klmo-knowledge}, optimizing \textit{fine-tuning} with knowledge-aware objectives~\cite{liang2024kagboostingllmsprofessional}, enhancing \textit{retrieval} through hybrid neural-symbolic architectures~\cite{izacard2020leveraging}, and refining \textit{prompt-based reasoning} via KG-guided subgraph extraction and logic-aware chained inference~\cite{choudhary2023complex}. 

\par Recent studies have begun exploring the \review{Collaboration Between Human, LLMs, and KG}, such as KNOWNET\cite{yan2024knownet}, which combines LLMs with KGs to enhance the quality and efficiency of human-AI interaction through structured knowledge validation and iterative query refinement. While integrating LLMs with KGs can enhance knowledge coverage and reasoning, existing approaches often fail to apply LLMs' reasoning capabilities to KGs directly and underestimate the role of human experts in decision-making, limiting collaborative efficacy. To address this, we propose a hypothesis validation framework that harnesses the complementary strengths of LLMs, KGs, and human experts. By leveraging each component's strengths and mitigating limitations, our framework promotes transparent, controllable collaboration, enhancing hypothesis validation and discovery outcomes.

\section{Formative Study}
\par In this study, we aimed to understand researchers' challenges, expectations, and requirements for collaborating with AI across various domains comprehensively. To achieve this, semi-structured interviews \review{(see Appendix E for details)} with institutional IRB approval were conducted with two researchers specializing in cancer therapy (\textbf{E1}-\textbf{E2}), and two researchers focusing on drug research (\textbf{E3}-\textbf{E4}) (Mean Age = $39.75$, SD = $5.4$, $2$ males, $2$ females). Among them, \textbf{E1}-\textbf{E2} specialize in screening SL pairs using the \textit{CRISPR/Cas9} technique, focusing on thyroid cancer (\textbf{E1}) and breast cancer (\textbf{E2}), respectively. \textbf{E3}-\textbf{E4} are engaged in drug research, with \textbf{E3} focusing on drug repurposing, while \textbf{E4} specializes in discovering new drug targets through pharmacogenomics. Each participant has significant research experience in their respective fields and expertise in AI-assisted prediction. Through inductive coding and thematic analysis~\cite{clarke2017thematic}, we extracted valuable insights into the challenges faced by domain experts and summarized the design requirements. Each interview session lasted approximately $45$ minutes.

\subsection{Experts' Conventional Practices}
\par \textbf{E3}, specializing in drug repurposing, outlined their conventional approach to mechanistic investigations \cite{hasin2017multi}. The process begins with selecting a research domain, followed by an extensive review of background literature to establish foundational knowledge. Combining this information and their domain expertise, they formulate initial hypotheses through logical reasoning, which are then tested experimentally via methods like high-throughput sequencing. However, \textbf{E3} highlighted limitations in this traditional workflow: \textit{Research directions are often restricted to familiar fields to reduce uncertainty, and hypothesis development requires laborious literature reviews to balance novelty with plausibility.} This is particularly challenging in less familiar domains, where limited expertise can undermine confidence in hypothesis design. Additionally, over-relying on trial-and-error experiments without theoretical grounding wastes resources and limits progress.

\par \textbf{E1} noted that the similar workflows are also applied to SL mechanistic studies, but with one critical distinction: Wet-lab experiments in SL research take around six months to complete, dramatically increasing the cost of trial-and-error. This necessitates rigorous hypothesis evaluation and meticulous experimental planning prior to testing. To facilitate this process, researchers now employ GNN models to predict potential SL pairs. Predictions are filtered using two criteria: (1) the interpretability of interpretative paths\footnote{\small{The sequence of entities and edges connecting a head entity (e.g., a gene) to a predicted SL partner via biologically meaningful relationships.}} and (2) the model's prediction confidence scores. While this approach reduces reliance on purely familiarity-driven hypothesis, significant challenges remain: \textit{Human intervention is still required to evaluate predictions, and generating truly novel hypotheses continues to demand substantial creativity and domain insight}. Given the large volume of predicted results, the experts have explored the use of LLMs for direct reasoning and filtering~\cite{10.1007/978-981-97-9434-8_33 ,schimanski-etal-2024-towards}. However, when attempting to integrate LLMs into their research, they found it challenging to effectively incorporate local data (e.g., KG). Moreover, their practical use revealed that LLMs often struggle to maintain consistency and coherence in handling complex hypotheses.


\subsection{Experts' Concerns and \review{Expectations}}
\par Initially, domain experts expressed concerns that structured knowledge---often represented as triplets, the basic units of KGs describing relationships between entities (e.g., \textit{Primaquine}-\textit{Affects}-\textit{IKBKG})---lacks commonsense details and explanatory text. They also noted that the granularity of edge relationships is sometimes too coarse, making it difficult to interpret different paths. As one expert pointed out: ``\textit{... some GNNs do give us interpretative paths, but honestly, the information these paths contain is pretty limited. You know, even when different groups of entities are linked by the same label, the actual meaning can be very different. [Point at CYLC1 and DAB1] They are both labeled as being involved in cell differentiation, but I'd bet their actual roles are pretty different. All these missing details make interpretative paths hard to understand and, honestly, it just makes me not trust them as much.}'' This reveals the first challenge: \textbf{C1. Insufficient detail and inadequate granularity in interpretative paths.} \review{To address the issue of limited detail and insufficient granularity in interpretative paths \textbf{[C1]}, it is crucial to provide reliable and comprehensive information. This leads to our first design requirement: \textbf{DR1. Enriching interpretative paths with reliable and sufficient detail}}.


\par Experts have also expressed concerns about the potential inadequacy of relevant information for hypothesis construction. When formulating hypotheses, domain experts often require supplementary knowledge to integrate diverse information and generate novel insights. Acquiring such knowledge typically involves extensive literature reviews, which can be burdensome and prone to overlooking critical insights. As \textbf{E1} emphasized, ``\textit{... When aiming to construct novel hypotheses, relying solely on our existing knowledge is often insufficient. We need to gather as much relevant information as possible to inspire new ideas and facilitate reasoning. However, this process is time-consuming and mentally demanding...Sometimes, extensive reading may not always help us grasp key points, and prolonged reading sessions can be distracting.}'' This highlights the second challenge: \textbf{C2. Inefficiency in acquiring relevant knowledge for novel hypothesis construction.} \review{Building on this, experts further emphasized the need for more effective methods, such as KG-based information retrieval, semantic search, and recommendation to aid information retrieval, ultimately facilitating hypothesis construction. This leads to the second design requirement: \textbf{DR2. Providing efficient methods to obtain relevant information for hypothesis construction}.}

 
\par Managing large-scale model predictions presents another significant challenge. The most reliable method currently involves domain experts manually analyzing and filtering predictions to identify novel research focuses or by some visualization methods~\cite{shi2022medchemlens}. However, as the volume of predictions increases, this manual or visual-assisted approach becomes unsustainable, placing a heavy burden on experts and potentially affecting their judgment. \textbf{E2} illustrated this difficulty: ``\textit{Prediction models are pretty impressive---they do get some things right. But with so many results, it’s like going from finding a needle in the ocean to a needle in a pond. Trying to sort through everything on our own without a clear direction is still overwhelming and just not realistic.}''  This highlights the third challenge: \textbf{C3. Impracticality of manually filtering and analyzing large-scale predictions}. \review{In light of the above challenge, \textbf{E1} highlighted, ``\textit{It's practically infeasible and really limits how much we can focus. Rule-based methods? Yeah, they can simplify things, but they'll definitely miss a lot of results that don't fit perfectly. We really need something smarter and more intuitive to quickly zero in on the results that actually matter.}'' This insight leads us to our third design requirement: \textbf{DR3. Streamlining and optimizing evaluation and insight extraction in massive predictions}.}

\par Experts also noted that while existing tools can help uncover simple patterns, they often reinforce known results rather than discovering novel insights. As \textbf{E2} explained: ``\textit{...Some tools can help us with simple patterns, but they just find results that look like what we already know. The problem is, if a model predicts something different---even if it's right---we might not even notice. That keeps us stuck in a loop, always going for the easy, obvious predictions. Suppose we really want to dig deeper and uncover complex mechanisms, then it's not just about finding similar predictions...We need better ways to build logical connections with hypotheses through observations, refine them, and search for patterns that actually lead us to new insights that we might've missed before.}'' This reveals the fourth challenge: \textbf{C4. Lack of tools to construct hypothesis chains for exploring complex mechanisms}. \review{As \textbf{E3} noted, ``\textit{When we're building a hypothesis, every step needs to make sense. If one part doesn't, the [whole] thing falls apart. That's why we need something to help us catch [any] flaws. And expanding the hypothesis is tricky too---it can feel pretty random and really depends on what we already know. A tool that suggests reasonable hypotheses and explains them? Yeah, I'd definitely be interested in trying that.}'' This leads to the fourth design requirement: \textbf{DR4. Supporting interactive construction and continuous optimization of hypothesis chains}.}

\par Even with tools for constructing hypothesis chains, experts found it difficult to retrieve relevant predictions that align with their hypotheses. They emphasized the need for better integration between model predictions and KG to refine hypotheses. \textbf{E4} explained, ``\textit{If we construct a hypothesis chain based on certain observations and our expertise but are unable to retrieve similar predictions or relevant evidence from the KG, it does not seem to help us come up with more solid hypotheses or further gain new insights}''. This highlights the fifth challenge: \textbf{C5. Absence of retrieval methods for hypothesis-aligned predictions}. As \textbf{E2} and \textbf{E4} noted, ``\textit{It’s exciting to find known paths or predictions closely aligning with our hypotheses.''} \review{However, designing search logic or rules for each hypothesis proves difficult, they added, ``\textit{and on top of that, it is impossible to experimentally validate every emerging hypothesis.''} This highlights the fifth design requirement: \textbf{DR5. Providing effective retrieval methods for hypothesis-aligned predictions}.}

\par Experts have also raised concerns about the information flow between LLMs, KGs, and domain experts, particularly based on their prior experiences with LLM integration. Despite efforts to bridge these components, significant challenges and \review{barriers} remain in ensuring that information is effectively transmitted and utilized. As \textbf{E3} noted after the discussion, ``\textit{I’m hoping LLMs and KGs can better share information with each other. Right now, when I use an LLM with a KG, the LLM often relies too much on external knowledge and doesn’t fully use the KG information unless I explicitly highlight the details. At the same time, the KG struggles to recognize entities when the LLM describes them differently, making the whole interaction awkward and fragmented. I want to make this smoother, improve how they understand each other, and help them combine their strengths to reach stronger conclusions.}'' This highlights the last challenge: \textbf{C6. Information transmission barriers between LLMs, KGs, and domain experts}. \review{Accordingly, we introduce the last design requirement: \textbf{DR6. Ensuring seamless and lossless transmission of information between LLMs, KGs, and domain experts}. This requirement aims to consistently maintain the quality of information, ensuring that when accurate information is transmitted between different parties, its completeness and precision are always preserved. Reflecting the perspectives of \textbf{E3}, ``\textit{Sure, we obviously want the info to be reliable when it's passed between the LLM, KG, and us. If there's any mix-up or something gets left out, it's definitely gonna mess with our judgments. I think this is really the key to making sure we can work together smoothly.}''}

\section{HypoChainer}

\par To ensure a seamless flow of information among different parties and enable experts to efficiently explore the system, the pipeline has been meticulously designed and iteratively refined. It comprises three key modules: \textbf{Contextual Exploration}, \textbf{Hypothesis Construction}, and \textbf{Validation Selection}, aligning with the conventional workflow while improving overall coherence, rationality, and efficiency (\cref{fig:pipeline}) (\textbf{DR6}). This section first provides a detailed overview of the backend algorithms implemented, followed by an introduction to the system's visual design. Finally, a walkthrough case study is presented to offer a concrete and comprehensible introduction to the entire pipeline.

\subsection{Data and Backend Engine}
\par We provide an overview of the data utilized in our approach. To assess the effectiveness and generalizability of our method, we employ two types of biological data: \textit{Drug Repurposing} and \textit{Cancer Research}.

\par \textbf{Drug Repurposing.} A large-scale biomedical knowledge graph~\cite{ma2023kgml} was utilized, derived from \textit{RTX-KG2c (v2.7.3)}, which integrates data from $70$ public sources, with $6.4$ million entities and $39.3$ million edges. The graph is standardized using the Biolink model~\cite{https://doi.org/10.1111/cts.13302}. To tailor it for drug repurposing, the dataset was refined into a streamlined graph with $3,659,165$ entities and $18,291,237$ edges. Key data sources include \textit{MyChem}, \textit{SemMedDB}, \textit{NDF-RT}, and \textit{RepoDB}, providing both positive (indications) and negative (contraindications/no-effect) samples. Furthermore, $472$ drug-disease pairs are leveraged from \textit{DrugMechDB} for external validation, enhancing the dataset's reliability and applicability. 

\textit{KGML-xDTD}~\cite{ma2023kgml} is a drug repurposing prediction and mechanism of action (MOA) inference model, integrating graph-based learning with reinforcement learning. The module formulates the task as a link prediction problem, which employs \textit{GraphSAGE}, to generate node embeddings by leveraging structural and attribute information from \textit{PubMedBERT}. The MOA module identifies biologically plausible MOA paths using an adversarial actor-critic reinforcement model, which is trained using curated demonstration paths. \textit{KGML-xDTD} enhances prediction (Appendix Tab. 1) while maintaining interpretability.

\par \textbf{Cancer Research.} This dataset~\cite{zhang2023kr4sl} focuses on SL relationships, where the simultaneous inactivation of a gene pair leads to the death of a specific cancer cell. The cancer research knowledge graph is constructed using data from two primary repositories: \textit{SynLethDB}, which catalogs validated gene SL interactions, and \textit{ProteinKG25}, a biomedical knowledge dataset containing information on gene functions, pathways, and biological processes. The resulting KG comprises $42,547$ entities, $33$ edge types, and a total of $396,619$ triplets.

For SL gene pair prediction, we utilize the \textit{KR4SL} model~\cite{zhang2023kr4sl}, which follows an encoder-decoder architecture and employs a GNN-based approach with a heterogeneous KG. The model predicts SL pairs by tracing relational paths, assigning weights to these paths to capture the strength of SL interactions and uncover potential relationships between unconnected genes. Candidates are ranked based on their likelihood of forming an SL relationship. Each prediction is accompanied by a three-hop interpretative path. The model achieves a precision of $59\%$ (Appendix Tab. 2), outperforming comparable models.

\subsection{Frontend Interface}
\par Working alongside domain experts, we have designed a frontend interface to support scientific discovery workflows through a set of dedicated modules, namely \textit{Control Panel}, \textit{Embedding View}, \textit{Chatbot}, \textit{Prediction View}, \textit{Hypothesis View}, \textit{Chain View}, and \textit{Retrieval View}.

\textbf{Control Panel.} The \textit{Control Panel} (\cref{fig:teaser}-\specialcircledgsingle{A}) includes three search boxes to streamline prediction selection across hierarchical levels. The \textit{Category} search box displays categories to assist in selecting specific types of \textit{Head Entities} or narrowing the scope of subsequent searches, while the \textit{Head} search box leverages an auto-complete function for direct \textit{Head Entity} queries. Upon selecting a \textit{Head Entity}, the \textit{Tail Entity} table positioned below the search boxes automatically populates with the top $50$ predicted \textit{Tail Entities}, sorted in ascending order by rank and annotated with their \textit{Name}, \textit{Score}, and \textit{Rank}. However, given that not all datasets contain predefined categories, users can define new categories encompassing relevant entities through the \textit{ChatBot} during the \textbf{Contextual Exploration} phase. Each query is logged in the \textit{Category} search box for convenient reuse in subsequent explorations.

\textbf{Embedding View.} The \textit{Embedding View} (\cref{fig:teaser}-\specialcircledgsingle{C}) displays cluster summaries derived from entity features, with methodologies tailored to distinct datasets. For instance, dimensionality reduction for gene data incorporates gene descriptions, nucleotide sequences, and other pertinent information, whereas drug data reduction utilizes attributes like drug descriptions, indications, and mechanisms of action through UMAP\cite{McInnes2018UMAPUM} (Comparison in Appendix). This aids domain experts in analyzing prediction patterns, offering an intuitive framework to identify research focuses within large-scale predictions (\textbf{DR3}). Additionally, the \textit{Embedding View} interacts with the \textit{ChatBot} during the \textbf{Contextual Exploration} phase: when the RAG suggests entities relevant to the query, \textit{Embedding View} highlights them for \textit{lasso} selection.

\textbf{ChatBot. }The \textit{ChatBot} (\cref{fig:teaser}-\specialcircledgsingle{B}) serves as the central hub of the system, orchestrating seamless interactions across all interface components by integrating reasoning models (\raisebox{-0.6ex}{\includegraphics[height=2.5ex]{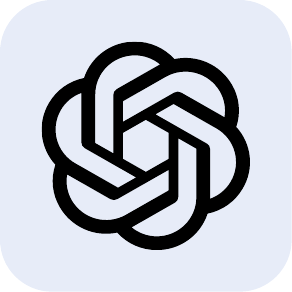}} LLM) and retrieval mechanisms, which is based on \textit{LightRAG}~\cite{guo2024lightrag} for its lower cost and faster response compared to \textit{GraphRAG}\cite{edge2024local} (\raisebox{-0.6ex} {\includegraphics[height=2.5ex]{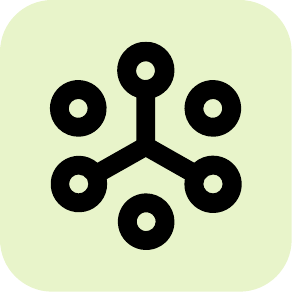}} RAG), to synchronize KG data, AI-generated insights, and expert input (\textbf{DR6}). The interface ensures intuitive usability through features such as a left \textit{history} section that logs and summarizes the dialogue, a bottom \textit{input box} with retrieval mode selector (\raisebox{-0.6ex} {\includegraphics[height=2.5ex]{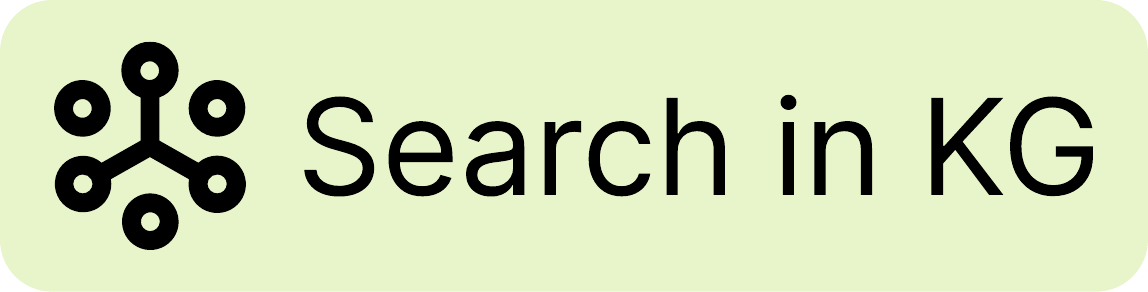}} \& \raisebox{-0.6ex} {\includegraphics[height=2.5ex]{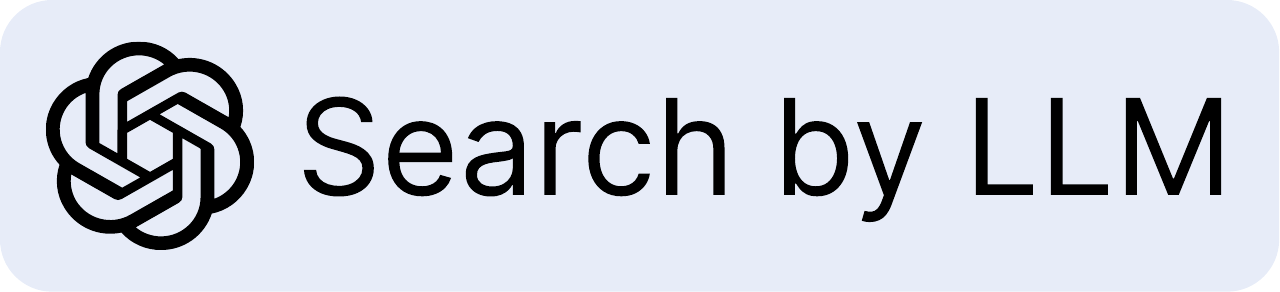}}), and upper-right \raisebox{-0.6ex}{\includegraphics[height=2.5ex]{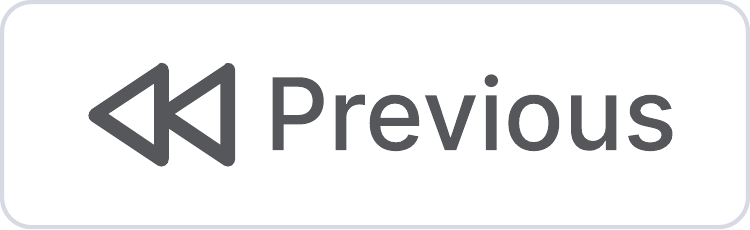}} \& \raisebox{-0.6ex} {\includegraphics[height=2.5ex]{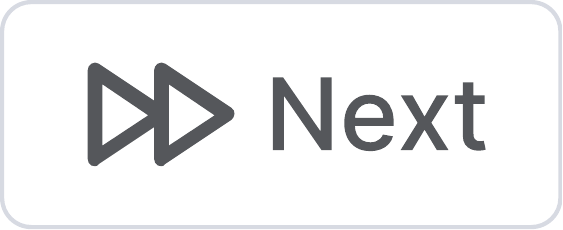}} buttons that activate phase-specific LLMs. For instance, during the \textit{Contextual Exploration}, RAG and the entity filtering and recommendation are invoked to ensure that the information comes solely from the local database, thus minimizing the risk of recommending entities not present in the KG or generating hallucinations. The chatbox dynamically adapts its color scheme to signal active retrieval modes, serving as a reminder to domain experts of the current model in use. Furthermore, the responses generated by the LLM include suggestions for further exploration, which are based on neighboring entities in the KG or relevant information related to the topic, and are displayed beneath each chatbox. Also, entities that appear in both the \textit{ChatBot} and the corresponding view are color-matched to ensure consistency, helping experts quickly locate and associate relevant entities.


\textbf{Prediction View. } The \textit{Prediction View} (\cref{fig:teaser}-\specialcircledgsingle{D}) is designed to empower users in distilling valuable insights from a substantial volume of predictions (\textbf{DR3}), enabling targeted identification of results aligned with their research objectives. Building on the foundational principle of \textit{LineUp}~\cite{Gratzl2013LineUp}, this view adopts a tabular visualization that balances domain experts' familiarity with advanced analytical capabilities. The design optimizes spatial efficiency by binding entities to their subsequent edge relationships, compressing horizontal layouts without sacrificing contextual continuity. To streamline exploration, the system provides two key features: auto-completion-enabled filtering, which supports multi-hop queries across entities and edge relations, and dynamic highlighting mechanisms that enhance analytical precision. When users hover over entities, the system highlights the full corresponding interpretative path (\cref{fig:teaser}-\circled{6}), aiding the exploration of connection patterns. \review{A persistent anchoring bar at the bottom of the view displays the hovered path and fixes the selected one, reducing visual tracking effort and supporting path comparison.} \review{Once a hypothesis chain is submitted for analysis in the \textit{Chain view}}, hypothesis-aligned predictions are marked by \textcolor{red}{\ding{72}} (\cref{fig:teaser}-\circled{10}) to enable users to quantify the consistency between theoretical assumptions and empirical outcomes. Additionally, experts are able to perform single-column sorting based on path scores and edge weights. These features collectively enable domain experts to iteratively refine insights---from macro-level pattern discovery to granular path analysis---by systematically prioritizing paths through sortable confidence scores, cross-validating hypotheses against prediction, and exporting candidate paths for validation.



\textbf{Hypothesis View.} The \textit{Hypothesis View} (\cref{fig:teaser}-\specialcircledgsingle{E}) presents KG entities aligned with a selected prediction path (\textbf{DR2}), enrich structured information (\textbf{DR1}), and assist domain experts in hypothesis refinement through iterative optimization (\textbf{DR4}). The view dynamically retrieves entities through RAG and organize directly connected 1-hop neighbors in KG (\review{Appendix Fig. 1}-\alterrectanglered{1}) and hypothesis-aligned entities (\review{Appendix Fig. 1}-\alterrectangleblue{2}) retrieved from the generated hypotheses and the subsequent iterative refined hypotheses in a hierarchical architecture. To optimize layout clarification and scalability for large biomedical datasets, entities within each layer are arranged using a Voronoi treemap~\cite{balzer2005Voronoi}, \review{a technique proven effective in reducing visual clutter while preserving spatial efficiency, making it particularly suitable for visualizing large-scale knowledge graphs~\cite{jiang2024slinterpreter}.} Entities of the same type are clustered within layers to highlight structural patterns and simplify analysis. \review{When users hover over a node in the Voronoi treemap, the node and its connected edges are highlighted in red, directing attention to relevant relationships.}

\par The view further dynamically establishes plausible KG-derived links between entities across adjacent layers. This approach enables domain experts to progressively and intuitively analyze potential connections between 1-hop entities on the predicted path and hypothesis-aligned semantic matches, fostering a structured understanding of both KG proximity and semantic relevance. By visualizing potential connection patterns, experts can assess whether these connections align with similar underlying hypotheses, thereby refining validation processes and guiding the construction of logically coherent hypothesis chains.

\textbf{Chain View.} The \textit{Chain View} supports domain experts in systematically constructing, previewing, and refining hypotheses (\textbf{DR4}). It facilitates the iterative analysis and optimization of individual hypotheses through integration with LLMs, enabling users to chain hypotheses into a cohesive structure for subsequent retrieval and validation.

\par The \textit{Input Area} (\cref{fig:teaser}-\specialcircledg{F}{1}) enables hypothesis chain construction and review. Each hypothesis node features a text area above for hypothesis description, with additional fields below to define relationships between entities. Clicking the lower right \raisebox{-0.6ex}{\includegraphics[height=2.5ex]{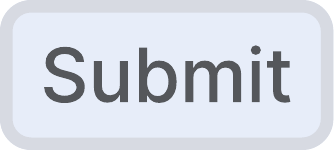}} button triggers targeted RAG retrieval to validate the hypothesis against the KG, displaying relevant entities in the \textit{Entity Preview} (\cref{fig:teaser}-\specialcircledg{F}{2}), including entity names, types, and descriptions of how they align with the hypothesis, \review{ordered by their degree of alignment, which is judged by the RAG through a systematically designed prompt.} This preview facilitates rapid hypothesis evaluation and refinement without resource-intensive full-scale RAG retrievals. At the top-right corner, the \raisebox{-0.6ex} {\includegraphics[height=2.5ex]{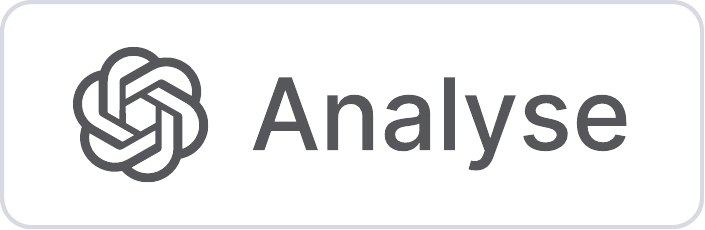}} button leverages LLM capabilities to evaluate the hypothesis chain for logical coherence, identifying potential inconsistencies or optimization opportunities. Once validated, the \raisebox{-0.6ex} {\includegraphics[height=2.5ex]{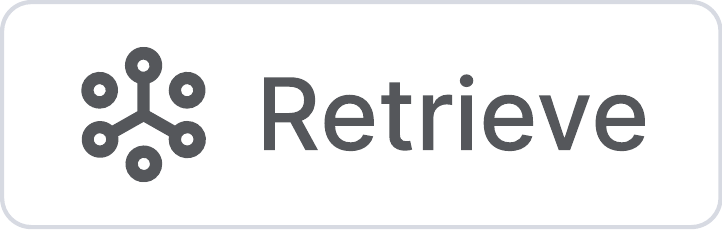}} button initiates the formal retrieval process for downstream tasks. To clarify functionality, the \raisebox{-0.6ex} {\includegraphics[height=2.5ex]{figs/retrieve.pdf}} button (linked to KG-grounded RAG retrieval) and the \raisebox{-0.6ex} {\includegraphics[height=2.5ex]{figs/analyse.pdf}} button (LLM-driven reasoning) are differentiated through the icons in \textit{ChatBot}.

\textbf{Retrieval View.} The \textit{Retrieval View} provides an overview of all hypothesis chain retrievals and corresponding results. It highlights the quantities of retrieval outcomes aligned with the hypothesis chain at varying matching levels, alongside detailed retrieved results (\textbf{DR5}).

\par
\begin{figure}[h]
\centering
\vspace{-3mm}
\includegraphics[width=\linewidth]{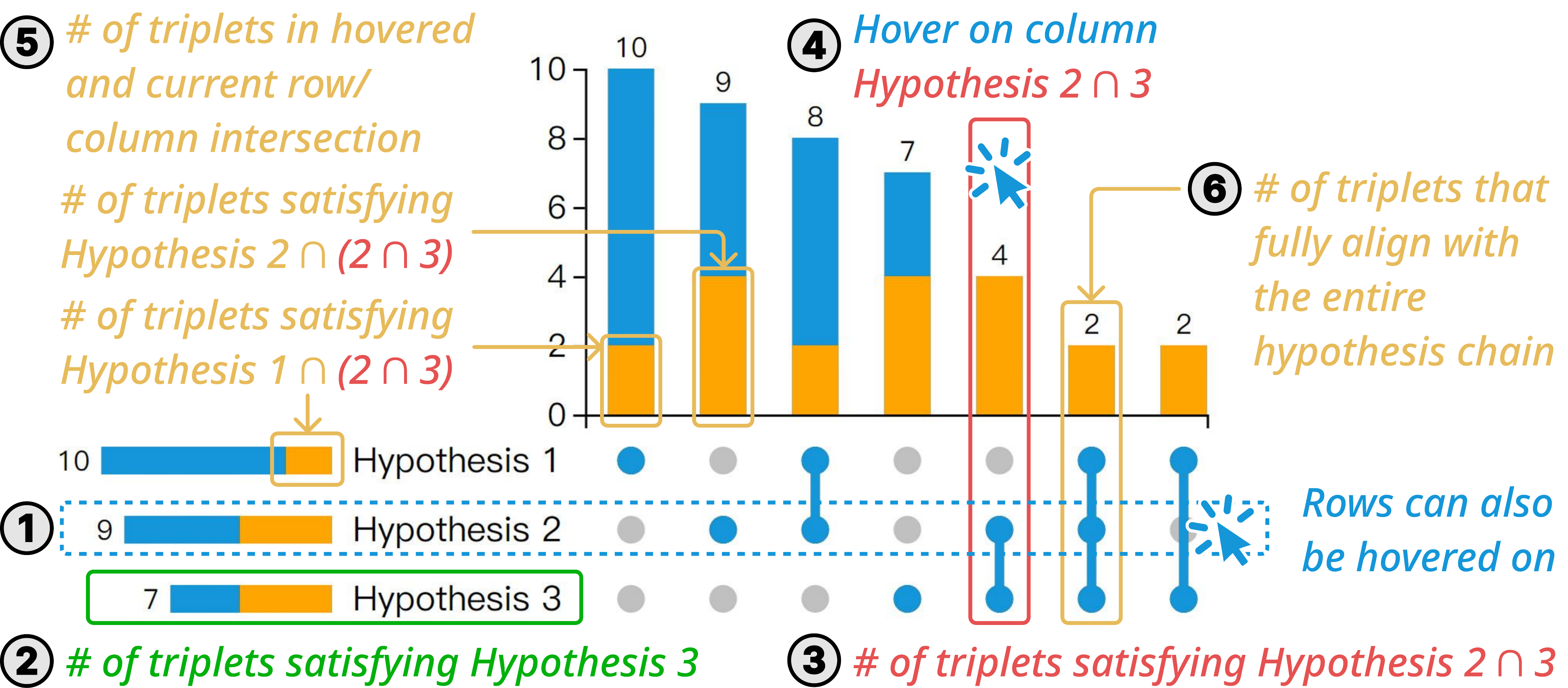}
\vspace{-6mm}
\caption{UpSet Plot: Visualization of entity triplets alignment organized by intersections and inclusion relationships among hypotheses.}
\label{fig:upset}
\vspace{-3mm}
\end{figure}

 The \textit{Retrieval List} (\cref{fig:teaser}-\specialcircledg{G}{1}) catalogs hypothesis chains and their retrieval outcomes, organized into collapsible records. Below, the \textit{UpSet Plot} (\cref{fig:teaser}-\specialcircledg{G}{2})~\cite{Lex2014upset} visualizes the degree to which retrieval results for the currently selected hypothesis chain satisfy the hypothesis, assisting experts in identifying which hypotheses require refinement or commonly appear in predictions. Specifically, the \textit{UpSet Plot} includes three rows (\cref{fig:upset}-\protect\circled{1}) representing the three hypotheses within the chain, with an adjacent bar chart (\cref{fig:upset}-\protect\circled{2}) showing the count of triplets satisfying individual hypotheses. A central dot matrix and bar chart (\cref{fig:upset}-\protect\circled{3}) represent combinatorial hypothesis satisfaction. For example, the fifth column highlights triplets fulfilling both Hypothesis 2 and 3 in the hypothesis chain. Hovering over a row or column (\cref{fig:upset}-\protect\circled{4}) dynamically highlights intersecting sets and displays their proportional contributions within the corresponding hypothesis (\cref{fig:upset}-\protect\circled{5}). Clicking bars filters the \textit{Retrieval List} accordingly, with intersections indicated at the upper-right corner (\cref{fig:teaser}-\circled{15}). The \raisebox{-0.6ex} {\includegraphics[height=2.5ex]{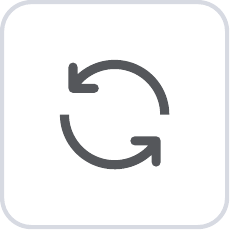}} button resets the view to the default display of complete retrieval results.

\subsection{Pipeline Walkthrough}
\par To clarify the pipeline's structure, we walked through an SL prediction example together with \textbf{E1} and \textbf{E2}, systematically demonstrating how \textbf{Contextual Exploration}, \textbf{Hypothesis Generation}, and \textbf{Validation Selection} are applied in sequence, with detailed steps and explanations.

\begin{figure*}[h]
\centering
\vspace{-3mm}
\includegraphics[width=\textwidth]{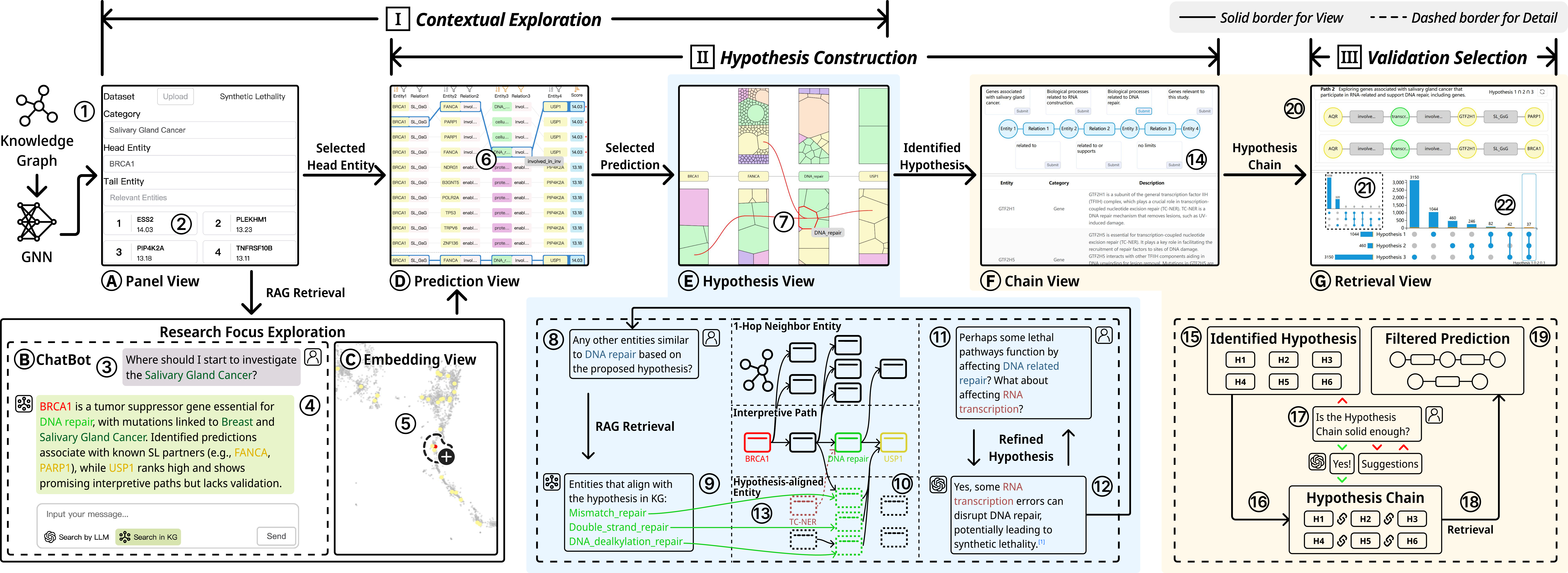}
\vspace{-6mm}
\caption{The pipeline comprises three main components: \protect\whiterectangle{I} \textbf{Contextual Exploration}, \protect\whiterectangle{II} \textbf{Hypothesis Construction}, and \protect\whiterectangle{III} \textbf{Validation Selection}.}
\label{fig:pipeline-backend}
\vspace{-6mm}
\end{figure*}

\par \textbf{Contextual Exploration.} At this stage (\cref{fig:pipeline-backend}-\whiterectangle{I}), the system provides experts with comprehensive and relevant information about their selected research focuses. It explains predictions with supplementary information on edge relationship granularity and integrates structured KG knowledge to support predictions. This structured information facilitates further mechanism exploration, potentially revealing the underlying mechanisms behind the predictions.

\par The process begins with training the GNN model on full-scale KG (\cref{fig:pipeline-backend}-\caserectangle{1}), generating predictions along with interpretative paths. To assist experts in identifying research focuses matching their expertise and interest, while also filtering the most relevant predictions from vast results (\textbf{DR3}) (\cref{fig:pipeline-backend}-\caserectangle{2}), the system incorporates a RAG-based retrieval framework with two modes: 1) Online LLM retrieval tailored to the biomedical domain. 2) RAG retrieval exclusively using local knowledge to prevent external data interference. Domain experts can engage with the RAG system (\cref{fig:pipeline-backend}-\specialcircledgsingle{B}) by formulating queries based on their interests. For example, \textbf{E1}, an expert in breast cancer SL mechanisms, was interested in exploring SL in \textit{salivary gland cancer} to identify potential mechanisms or patterns similar to \textit{breast cancer}. To initiate this research, he queried RAG: ``\textit{I would like to conduct research on salivary gland cancer, and I am an expert in breast cancer. Could you suggest some relevant predictions to facilitate the commencement of my study?}'' (\cref{fig:pipeline-backend}-\caserectangle{3}) The system then returned \textbf{recommendations} for each research focus with the most relevant entities (\cref{fig:pipeline-backend}-\caserectangle{4}), highlighted in the \textit{Embedding View} (\cref{fig:pipeline-backend}-\specialcircledgsingle{C}). Domain experts can refine their selection by integrating insights from RAG recommendations and the \textit{Embedding View} using lasso selections (\cref{fig:pipeline-backend}-\caserectangle{5}). As exemplified in the case, RAG first performed reasoning and local retrieval to generate a list of recommended genes. Among these, RAG recommended the gene \textit{BRCA1} due to its association with both \textit{salivary gland cancer} and \textit{breast cancer}. Despite extensive focus on \textit{BRCA1}, numerous unverified predictions persisted. Consequently, \textbf{E1} implemented a lasso selection on the cluster containing \textit{BRCA1}, subsequently filtering \textit{BRCA1} in the \textit{Prediction View} (\cref{fig:pipeline-backend}-\specialcircledgsingle{D}). Noticing many top-ranked predictions relied primarily on the \textit{sl\_gsg} relationship, which he considered biologically less meaningful, \textbf{E1} filtered out predictions whose interpretative paths consisted exclusively of \textit{sl\_gsg} connections. After that, he noticed that the prediction involving \textit{BRCA1} and \textit{USP1}, which was originally ranked $25^{\text{th}}$
, had moved up to the $2^{\text{nd}}$ position. Additionally, \textbf{E1} noticed experimentally validated genes \textit{FANCA} and \textit{PARP1} connected through two paths containing \textit{DNA\_repair}. Consequently, he identified \textit{USP1} as a valuable prediction for further validation, completing research focus selection in the \textit{Contextual Exploration} phase.
 
\par Once a research focus is chosen, experts can explore specific prediction paths (\cref{fig:pipeline-backend}-\caserectangle{6}) in the \textit{Hypothesis View} (\cref{fig:pipeline-backend}-\specialcircledgsingle{E}). This view further displays selected paths along with their adjacent graph structures, offering contextual insights into the research topic (\cref{fig:pipeline-backend}-\caserectangle{7}), aiding experts in comprehending the interpretative KG path and exploring additional related information. For example, \textbf{E1} queried for additional details regarding \textit{DNA\_repair} in the interpretative path (\cref{fig:pipeline-backend}-\caserectangle{8}), obtaining a detailed explanation enhanced by external knowledge.
 
\par Then the LLM, with external information access, enriches interpretative paths by providing additional details in the \textit{ChatBot}, allowing experts to bridge gaps in structured data by integrating common knowledge and cutting-edge insights not yet present in the KG. Experts can validate this information and, if valuable, integrate it into KG using text-to-KG methods within RAG, ensuring seamless KG expansion. Iterative querying and exploration allow experts to progressively refine their understanding of the predictions and the associated KG knowledge, laying the groundwork for subsequent hypothesis construction.

\par \textbf{Hypothesis Construction.} With prediction paths selected, domain experts require a solid basis to formulate hypotheses. Hence, the \textit{Hypothesis View} leverages the LLM to generate an initial hypothesis regarding the underlying mechanisms. At this stage (\cref{fig:pipeline-backend}-\whiterectangle{II}), the \textit{Hypothesis View} presents entities retrieved via RAG that align with the given hypothesis (\cref{fig:pipeline-backend}-\caserectangle{8}, \caserectangle{9}). Additionally, RAG paraphrases these descriptions, allowing domain experts to iteratively refine the reasoning until fully satisfied, ensuring an accurate understanding~\cite{li2024linkq}. As demonstrated in the example, \textbf{E1} wished to explore further, so he shifted to the \textit{Hypothesis View}. The LLM generated a potential hypothesis in the \textit{ChatBot} that the target gene likely shares SL relationship with \textit{DNA\_repair}-related genes through interactions involving known SL genes. \textbf{E1} found this consistent with his domain knowledge and proposed that if \textit{DNA\_repair} could connect to different SL genes, similar entities might also yield similar predictions (\cref{fig:pipeline-backend}-\caserectangle{8}). Consequently, the hypothesis was refined to encompass analogous entities related to \textit{DNA\_repair}. Then, the RAG retrieved three closely related entities: \textit{Mismatch\_repair}, \textit{Double\_strand\_repair}, and \textit{DNA\_dealkylation\_repair} (\cref{fig:pipeline-backend}-\caserectangle{9}). \textbf{E1} then observed that these entities were also linked to many analogous predicted entities (\cref{fig:pipeline-backend}-\caserectangle{10}), thus strengthening his belief in the hypothesis.

 
\par By leveraging these aligned entities, experts can concretize their hypothesis, examine related entities, and explore potential connection patterns predicted by the model (\cref{fig:pipeline-backend}-\specialcircledgsingle{E}). This iterative process (\textbf{DR4}) deepens their insights, allowing them to refine and adjust hypotheses based on emerging findings (\cref{fig:pipeline-backend}-\caserectangle{11}, \caserectangle{12}). As the hypothesis evolves, the displayed entities update accordingly. 
 
\par Throughout the iterative refinement, the LLM using external information provides heuristic guidance (\cref{fig:pipeline-backend}-\caserectangle{12}) in the \textit{ChatBot}, helping domain experts assess hypothesis validity, identify potential inconsistencies, and enhance logical coherence. Once satisfied with a constructed hypothesis (\cref{fig:pipeline-backend}-\caserectangle{15}), experts can integrate it into a \textit{hypothesis chain} within the \textit{Chain View} (\cref{fig:pipeline-backend}-\specialcircledgsingle{F}) for further retrieval.
 
\par A \textit{hypothesis chain} links multiple hypotheses, forming a structure similar to a triplet (\cref{fig:pipeline-backend}-\caserectangle{16}). However, it extends beyond simple entity-relationship pairs by incorporating textual descriptions to facilitate communication and encoding of hypotheses. Experts can continually refine existing hypotheses or develop new ones with LLM-generated suggestions (\cref{fig:pipeline-backend}-\caserectangle{17}), \review{With each submission of hypothesis analysis, aligned predictions are marked by \textcolor{red}{\ding{72}} in the \textit{prediction view}, helping experts refine the chain.} By constructing complex chains, they move beyond merely summarizing patterns or drawing simple analogies from model predictions. Instead, experts integrate expertise and insights into a flexible and targeted \textbf{Validation Selection} process, enabling deeper exploration of intricate underlying mechanisms. 
 
\par As shown in the case, \textbf{E1} recalled that some RNA transcriptions are important for \textit{DNA\_repair}. To explore further, he queried LLM whether the proposed hypothesis held water (\cref{fig:pipeline-backend}-\caserectangle{11}). The LLM confirmed a strong association between RNA and \textit{DNA\_repair} (\cref{fig:pipeline-backend}-\caserectangle{12}). However, subsequent analysis using RAG revealed a particular relationship: \textit{Transcription-Coupled Nucleotide-Excision-Repair} (\textit{TC-NER}) (\cref{fig:pipeline-backend}-\caserectangle{13}), a pathway where RNA polymerase triggers \textit{DNA\_repair} during transcription. Surprisingly, this relationship wasn’t directly linked to \textit{DNA\_repair} or analogous entities in KG. This prompted \textbf{E1} to ask the LLM for clarification. The LLM confirmed this was indeed a scientifically valid connection. Further exploration showed this relationship was only associated with a specific subset of genes, devoid of any discernible connections to other entities. \textbf{E1} then incorporated the relationship into the KG through the text-to-KG integration within RAG and adjusted the hypothesis chain to (\cref{fig:pipeline-backend}-\caserectangle{14}): [genes associated with \textit{salivary gland cancer}] $\rightarrow$ [biological processes related to RNA construction] $\rightarrow$ [biological processes related to \textit{DNA\_repair}] $\rightarrow$ [genes relevant to this study] (\cref{fig:pipeline-backend}-\caserectangle{16}), querying the LLM to ascertain the reasonableness of this chain (\cref{fig:pipeline-backend}-\caserectangle{17}). After reasoning, the LLM recommended refining the relationship between RNA and \textit{DNA\_repair}, suggesting that the chain should be rephrased to indicate that RNA transcription is ``related to'' or ``supports'' \textit{DNA\_repair}-related entities, which is predicated on the observation that, in certain instances, the inactivation of specific genes actually stimulate RNA transcription, which might not result in an SL pair. \textbf{E1} considered this suggestion reasonable and refined the hypothesis chain accordingly.

\par \textbf{Validation Selection}. At this stage (\cref{fig:pipeline-backend}-\whiterectangle{III}), after domain experts have formulated coherent and well-reasoned hypotheses, these hypotheses are retrieved based on all the entities identified through RAG to ensure retrieval accuracy and comprehensiveness (\cref{fig:pipeline-backend}-\caserectangle{18}) and are then compared against predictions and the KG (\cref{fig:pipeline-backend}-\caserectangle{19},\caserectangle{20}) within the \textit{Retrieval View} (\cref{fig:pipeline-backend}-\specialcircledgsingle{G}). This process helps determine whether similar conclusions have been previously validated or if relevant paths exist within the predictions.
 
\par Particularly, the retrieval process is guided by the entities proposed by RAG in the hypothesis chain (\cref{fig:pipeline-backend}-\caserectangle{19}), with the retrieved predictions grouped based on their alignment with the hypothesis (\cref{fig:pipeline-backend}-\caserectangle{20}). Domain experts can then examine these results to identify candidates for further experimental validation. Additionally, they can leverage the \textit{UpSet Plot} (\cref{fig:teaser}-\specialcircledg{G}{2}) to detect potential inconsistencies in the hypothesis during retrieval, gaining insights for further refinement and optimization. As reflected in the case, following the construction of the hypothesis chain, \textbf{E1} conducted a targeted retrieval (\cref{fig:pipeline-backend}-\caserectangle{18}) based on the refined hypothesis chain. Analysis of the results revealed that while many predictions included RNA transcription entities as intermediates and linked to final outcomes via \textit{DNA\_repair}, none of them fully aligned with the intermediate hypothesis criteria (\cref{fig:pipeline-backend}-\caserectangle{21}). Drawing from prior observations that certain RNA processes in the KG were predominantly connected to genes, \textbf{E1} expanded the hypothesis to incorporate gene-centric biological processes related to \textit{DNA\_repair}. Subsequent retrievals demonstrated strong alignment between predictions and the revised hypothesis chain (\cref{fig:pipeline-backend}-\caserectangle{22}). Notably, the gene \textit{EP300}—highly associated with \textit{salivary gland cancer}—was connected to predicted entities \textit{CYP2C9} and \textit{RAD23B} through RNA-transcription-related entities \textit{transcription-coupled\_nucleotide-excision\_repair} and \textit{DNA\_repair}-related genes \textit{GTF2H1}, \textit{GTF2H4}, and \textit{GTF2H5}. This reinforced the hypothesis that \textit{DNA\_repair} mechanisms may exhibit SL correlations across disease contexts, with RNA transcription acting as a potential extension of these mechanisms. However, the current KG lacked robust representations of RNA transcriptions–\textit{DNA\_repair} relationships, underscoring the need for supplemental data integration. These findings revealed areas for improvement in the KG and provided a novel perspective for further research into mechanistic synergies.

\vspace{-1mm}
\section{Evaluation}
\vspace{-1mm}


\par We conducted a case study and a user study to evaluate the effectiveness, workflow, and usability of \textit{HypoChainer}.

\vspace{-1mm}

\subsection{Case Study: Drug Repurposing Exploration}
\par \textbf{E3} and \textbf{E4}, specialists in drug mechanisms and gene therapy, focus on repurposing antiepileptic drugs (e.g., those treating spasms) to address other diseases and evaluate novel therapies. Their goals are twofold: (1) identify new applications for existing drugs and (2) assess their potential in mitigating complex diseases. They followed the Co-discovery Learning Protocol~\cite{lim1997empirical}, with one author guiding the session, \textbf{E3} operating the system, and \textbf{E4} discussing insights in real time.

\par To initiate this process, \textbf{E3} uploaded a drug repurposing dataset (\cref{fig:teaser}-\circled{1}) and queried the RAG module: ``\textit{I am researching antiepileptic drug repurposing. Can you suggest potentially related diseases?}'' (\cref{fig:teaser}-\circled{2}) The system performed a multi-faceted analysis of disease relationships, identifying $5$ high-relevance candidates, including \textit{Huntington's disease}, \textit{Episodic ataxia type 5}, \textit{Parkinson's disease}, \textit{Photosensitive tonic-clonic seizures}, and \textit{Generalized tonic-clonic seizures}. These results were prioritized in both textual and \textit{Embedding View}, with \textit{Episodic ataxia type 5} flagged as a top recommendation (\cref{fig:teaser}-\circled{3}). While \textbf{E3} had prior expertise in \textit{Photosensitive tonic-clonic seizures}, he used the \textit{Embedding View}'s lasso tool to isolate the \textit{Episodic ataxia} cluster (\cref{fig:teaser}-\circled{4}), opting to explore RAG's novel suggestion. In the \textit{Prediction View}, \textbf{E3} observed frequent top-ranked associations with the \textit{CACNA1C} gene (\cref{fig:teaser}-\circled{5}), \review{\textbf{E4} noted that this is a known regulator of voltage-gated calcium channels and a common target in epilepsy-related drug mechanisms.} To deepen \textbf{E3}'s investigation, he selected the less familiar \textit{Episodic ataxia type 5} prediction (\cref{fig:teaser}-\circled{6}), leveraging the system to bridge knowledge gaps and validate hypotheses.

\par \textbf{Hypothesis Refinement Workflow}. In the \textit{Hypothesis View}, the \textit{ChatBot} generated an initial mechanistic hypothesis via the LLM: [Modulating agent alters ion channel function] $\rightarrow$ [Impacts] $\rightarrow$ [Gene regulatory pathways (affected by calcium channels)] $\rightarrow$ [Drives] $\rightarrow$ [Neuronal network dynamics] $\rightarrow$ [Manifests as] $\rightarrow$ [Episodic neural dysfunction] (\cref{fig:teaser}-\circled{7}). \textbf{E3} first validated this chain (\cref{fig:teaser}-\circled{8}) by confirming that retrieved entities aligned with the hypothesis and were supported by contextual KG evidence (\cref{fig:teaser}-\circled{9}). He observed that integrating existing KG edges revealed novel sub-paths (\cref{fig:teaser}-\circled{10}) and multiple connections between hypothesis-aligned entities and one-hop neighbors, reinforcing the chain's coherence. Cross-referencing the \textit{Prediction View}, he noted consistency between the hypothesis and most interpretative paths (\cref{fig:teaser}-\circled{11}). \textbf{E3} considered this as evidence that the hypothesis chain explains most repurposing predictions.

\par \textbf{Critical Insight and Hypothesis Revision}. However, \textbf{E3} noted a misalignment between the hypothesis and high-score predictions for \textit{Huntington's disease} (\cref{fig:teaser}-\circled{12}). By filtering and examining these predictions, he observed that the original hypothesis aligned only partially. \review{Guided by the LLM’s explanations, \textbf{E4}’s further analysis of the remaining predictions revealed that some drugs were indicated for depression, while others were linked to neurodegenerative diseases.} This finding led \textbf{E3} to infer that the model's predictions regarding \textit{Huntington's disease} primarily reflect the symptomatic alleviation effects rather than an influence on or delay of its underlying etiology. To reconcile this, \textbf{E3} refined the hypothesis chain to: [\textit{Drugs treating motor dysfunction, depression, or neurodegeneration}] $\rightarrow$ [\textit{related to}] $\rightarrow$ [\textit{Interacting genes or pathways}] $\rightarrow$ [\textit{Participated in}] $\rightarrow$ [\textit{Processes related to motor dysfunction, depression, or neurodegeneration}] $\rightarrow$ [\textit{Led to}] $\rightarrow$ [\textit{Diseases associated with} \textit{Huntington's disease}].

\begin{wrapfigure}{r}{0.3\columnwidth}
 \vspace{-4mm}
 \centering 
 \includegraphics[width=0.3\columnwidth]{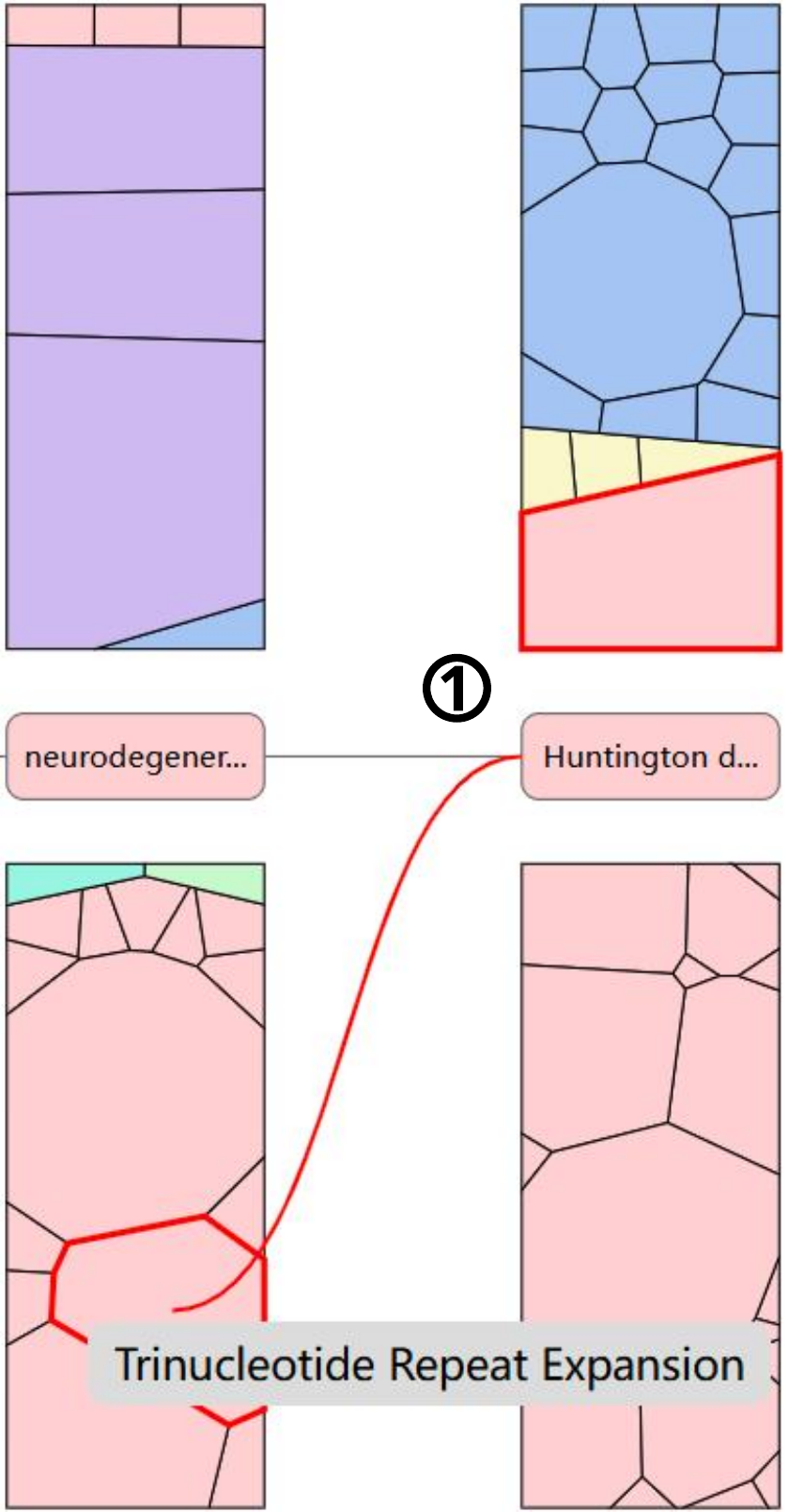}
 \vspace{-6mm}
 \caption{\protect\circled{1} \textbf{E3} noticed a link to \textit{Huntington's disease} from \textit{Trinucleotide Repeat Expansion}, a 1-hop entity of \textit{neurodegenerative disease}.}
 \label{fig:case2_2}
  \vspace{-3mm}
\end{wrapfigure}

\par \textbf{Hypothesis Challenge and Iterative Refinement}. During hypothesis validation, \textbf{E3} noted that the revised chain obscured the majority of the predictions. However, in the \textit{Hypothesis View}, he identified a link between \textit{Huntington's disease} and the \textit{Trinucleotide Repeat Expansion entity}—\review{a genetic mechanism associated with neurodegenerative disorders, as noted by \textbf{E4}} (\cref{fig:case2_2}-\circled{1}). Intrigued by its absence in the \textit{Prediction View}, \textbf{E3} queried the LLM, which explained that \textit{Trinucleotide Repeat Expansion} is a well-established genetic cause of \textit{Huntington's disease}, yet it did not surface in any \textit{Huntington}'s predictions. Intrigued by this discrepancy, \textbf{E3} revised the hypothesis chain to prioritize predictions involving \textit{Trinucleotide Repeat Expansion} and submitted the updated hypothesis chain via the \raisebox{-0.6ex} {\includegraphics[height=2.5ex]{figs/analyse.pdf}} for LLM validation. The LLM provided critical feedback: while \textit{Trinucleotide Repeat Expansion} refers to abnormal DNA sequence elongations that disrupt gene expression or protein function, its absence in predictions likely reflects the lack of direct therapeutics. However, the LLM emphasized indirect associations in literature, linking the entity to broader processes like \textit{DNA\_repair} and \textit{histone deacetylase (HDAC) regulation}.

\par \textbf{Insight-Driven Revision}. Guided by this feedback, \textbf{E3} refined the hypothesis to focus on \textit{HDAC}-related therapeutic targets. A subsequent retrieval (\cref{fig:teaser}-\circled{13}) identified therapies modulating \textit{HDAC} activity in diseases involving abnormal repeat expansions, such as \textit{Parkinson's disease}~\cite{kumar2022understanding} and \textit{Amyotrophic Lateral Sclerosis} (ALS) (\cref{fig:teaser}-\circled{16}). This alignment validated the revised chain (\cref{fig:teaser}-\circled{14},\circled{15}).

\par \textbf{Critical Discovery and Therapeutic Proposal}. Re-examining \textit{Huntington}'s predictions, \textbf{E3} uncovered a previously overlooked interpretative path: [\textit{Entinostat} $\rightarrow$ [decreases activity of] $\rightarrow$ [\textit{HDAC1} gene] $\rightarrow$ [interacts with] $\rightarrow$ [\textit{Histone H4}] $\rightarrow$ [gene associated with condition] $\rightarrow$ [\textit{Huntington's disease}]. This \textit{HDAC}-associated interpretative path aligned with literature suggesting \textit{HDAC}'s role in disease progression~\cite{hecklau2021effects}. Synthesizing these insights, \textbf{E3} and \textbf{E4} proposed the design of a novel \textit{cocktail therapy} combining three symptom-alleviating drugs with \textit{HDAC} inhibitors to potentially delay \textit{Huntington}'s progression. He further considered exploring \textit{HDAC}-targeted gene therapies as a complementary avenue for further research.

\review{\textbf{Takeaway Message.}  In the case study, \textbf{E3} and \textbf{E4} iteratively revised initial hypotheses as new insights emerged, reflecting a dynamic reasoning process. Although the final hypotheses diverged from the original, they represented logical extensions informed by system-driven exploration. For example, in repurposing drugs for \textit{Huntington's disease}, the expert initially focused on antiepileptic mechanisms for symptomatic relief. However, analysis of high-confidence predictions outside this scope revealed links to broader neurodegenerative pathways, prompting a shift toward a more integrative hypothesis. \textbf{E3} reported maintaining control over hypothesis development, with \textit{HypoChainer} offering timely support when exploration stalled—clarifying complex predictions, suggesting new directions, and highlighting supporting KG evidence. \textbf{E4} praised the system's alignment with conventional workflows, flexible entity categorization, and context-aware knowledge integration, which streamlined hypothesis refinement. They also noted that the system helped transcend initial cognitive frames and uncover unanticipated, yet scientifically valuable associations. This iterative refinement process underscores the system's potential to support deeper and more structured hypothesis generation in complex discovery tasks.}

\subsection{User Study}
\par We recruited 12 graduate students (Mean Age = $26.33$, SD = $1.93$; $6$ males, $6$ females) from bioinformatics or biomedical engineering backgrounds, including 6 PhD and 6 Master's students. Participants were randomly assigned to either the Baseline or \textit{HypoChainer} group, balanced by degree level (3 PhD and 3 Master's per group).

\par
\begin{figure}[h]
\centering
\vspace{-3mm}
\includegraphics[width=\linewidth]{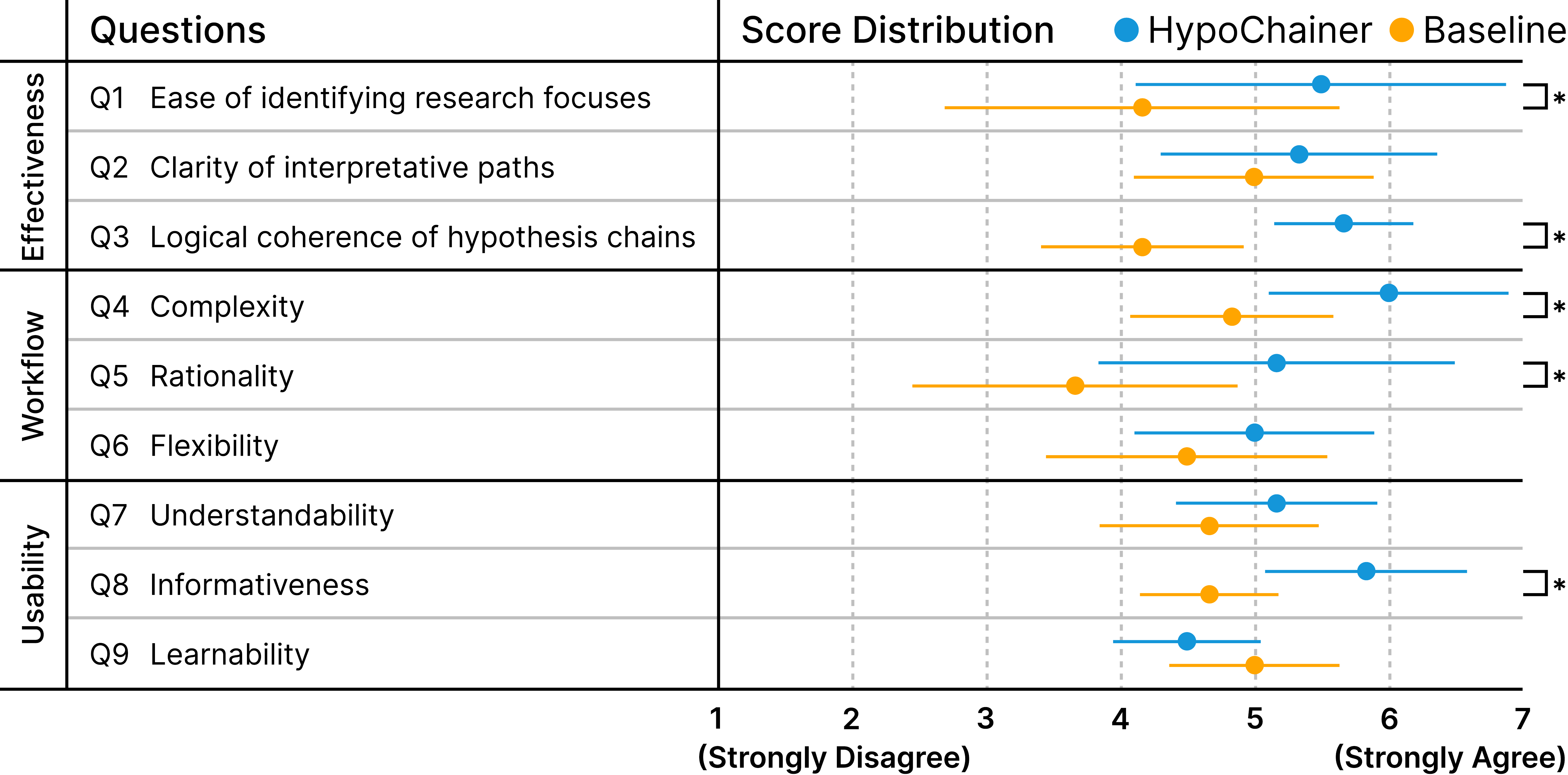}
\vspace{-6mm}
\caption{The questionnaire results of the two systems in terms of system effectiveness, workflow and usability ($* : p < 0.05$).}
\label{fig:questionnaire}
\vspace{-3mm}
\end{figure}

 Participants completed a drug repurposing task targeting \textit{Hemophilia B}. To ensure objective evaluation, all were screened to confirm no prior knowledge of the disease or its mechanisms. Five system-generated interpretative paths were selected based on partial alignment with DrugMechDB-curated mechanisms of actions (MOAs) for \textit{Eptacog Alfa} and \textit{Nonacog Alfa}. The Baseline system (Appendix Fig. 8) used an LLM-only setup, omitting the \textit{Hypothesis View}, to isolate the contributions of RAG and the view's design while maintaining the overall workflow. Both drugs were excluded from training data to better simulate real-world discovery conditions. Participants received a 30-minute training session on task goals, system usage, and evaluation criteria, followed by 90 minutes to explore predictions, formulate hypotheses, and submit results. The process was screen-recorded. A task was deemed successful if participants retrieved $\le$300 predictions and identified $\ge$3 of 5 reference predictions. A post-task questionnaire (\cref{fig:questionnaire}) adapted from the \textit{System Usability Scale}~\cite{sus} assessed perceived system effectiveness, workflow, and usability.

\par In the Baseline group, 2 PhD participants (33.3\%) completed the task within the time limit, compared to 4 in the \textit{HypoChainer} group (3 PhDs, 1 Master's; 66.7\%). PhD participants generally outperformed Master's students, likely due to stronger research backgrounds and faster system adaptation. \review{\textit{HypoChainer} showed clear advantages in \textit{Effectiveness}, with significantly higher scores for \textit{Ease of Identifying Research Directions} ($M=5.50$, $SD=1.38$, $p<0.05$) and \textit{Logical Coherence of Hypothesis Chains} ($M=5.67$, $SD=0.52$, $p<0.05$) compared to the Baseline ($M=4.17$, $SD=1.47$; $M=4.17$, $SD=0.75$). Both groups scored similarly on \textit{Clarity of Interpretative Paths} (\textit{HypoChainer}: $M=5.33$, $SD=1.03$; Baseline: $M=5.00$, $SD=0.89$).} \review{For \textit{Workflow}, \textit{HypoChainer} scored higher on \textit{Rationality} ($M=5.17$, $SD=1.33$, $p<0.05$) and \textit{Flexibility} ($M=5.00$, $SD=0.89$, $p<0.05$), supported by user feedback indicating fewer hallucinations. However, its integrated features—transitioning among LLM, RAG, and the Hypothesis View—were perceived as more complex ($M=6.00$, $SD=0.89$) than the Baseline ($M=4.83$, $SD=0.75$).} \review{In \textit{Usability}, \textit{HypoChainer} had a slightly lower score in \textit{Learnability} ($M=4.50$, $SD=0.55$) than the Baseline ($M=5.00$, $SD=0.63$), suggesting a steeper learning curve. Nevertheless, it received significantly higher ratings for \textit{Informativeness} ($M=5.83$, $SD=0.75$, $p<0.05$) and \textit{Understandability} ($M=5.17$, $SD=0.75$, $p<0.05$), due to its clear reasoning structure and entity-level explanations.} \review{Overall, \textit{HypoChainer} provided more effective, transparent, and insightful support for hypothesis generation, despite a slightly higher learning threshold.
}

\section{Discussion and Limitation}

\noindent\textbf{Lessons Learned.} During the evaluation of RAGs, we identified critical trade-offs between accuracy and computational cost. Tests revealed that response accuracy depends on the scale of the local KG and the entity retrieval limit per query. Increasing the retrieval limit from the default to more entities markedly improved accuracy but incurred higher token consumption and prolonged query times. This highlights the necessity of balancing query performance and cost in large-scale RAG-based workflows. Additionally, while model predictions occasionally diverged from the constructed hypothesis chains, domain experts emphasized that such discrepancies do not diminish the validity of the chains themselves. They highlighted that rigorously derived hypotheses—even without full alignment with retrieval results—retain significant values, particularly in uncovering novel insights.

\noindent\textbf{Generalization and Scalability.} \review{The hypothesis-driven architecture has demonstrated broad applicability across diverse research domains, particularly in scientific discovery tasks involving KG-based node and link prediction.} When integrated with RAG, the system offers a cost-effective and flexible alternative to domain-specific fine-tuning, enabling more efficient adaptation to new research areas. \review{Given the generalizability of RAG and the system's modular KG-based design, most components are transferable across domains, requiring only the substitution of domain-specific KGs and predictive models. Domain experts have also identified promising applications in areas such as protein structure and function prediction via interaction networks, as well as novel materials discovery. However, the system's performance is largely dependent on the quality of the underlying KGs and the effectiveness of the associated predictive models.} As advances in text-based KG extraction continue to improve accuracy and robustness, the system's cross-domain applicability is expected to expand further.


\noindent\textbf{Limitations.} \review{While RAG supports text-to-KG conversion, the process remains resource-intensive and prone to inaccuracies, especially in data-rich domains. To reduce risk, we restricted integration to small, verified updates using public KGs. However, the slow pace of KG updates compared to rapid scientific progress highlights the need for more accurate, automated extraction methods. Although RAG helps mitigate hallucinations by retrieving from traceable sources, such issues persist. Future improvements could include fact-verification modules—e.g., cross-referencing authoritative sources or applying confidence thresholds to flag uncertain outputs. Given the complexity of hypothesis-driven discovery, some interaction complexity is unavoidable, though we anticipate continued simplification as AI reliability advances.}


\section{Conclusion and Future Work}
\par This study introduces \textit{HypoChainer}, a collaborative framework that synergizes LLMs and KGs to advance hypothesis-driven scientific discovery. \textit{HypoChainer} provides three main functionalities: \textbf{Contextual Exploration}, \textbf{Hypothesis Construction} and \textbf{Validation Selection}. A case study and expert interviews demonstrate \textit{HypoChainer}'s capability to synthesize context-aware knowledge efficiently, construct and refine hypothesis chains systematically, and facilitate informed validation selections. Future work includes enhancing text-KG integration through more robust methods and accelerating knowledge discovery via more workflow automation, while preserving critical human oversight to ensure scientific rigor and actionable insights.

\acknowledgments{%
We gratefully acknowledge Dr. Jia Liu and Dr. Yifeng Yang from the Institute of Immunochemistry at ShanghaiTech University for their valuable collaboration, as well as the anonymous reviewers for their insightful feedback. This research was supported by the National Natural Science Foundation of China (No. 62372298), the ``AI Technologies for Accelerating Biopharmaceutical R\&D – School of Information Science and Technology'' (No. 2024X0203-902-01), the Shanghai Engineering Research Center of Intelligent Vision and Imaging, the Shanghai Frontiers Science Center of Human-centered Artificial Intelligence (ShangHAI), and the MoE Key Laboratory of Intelligent Perception and Human-Machine Collaboration (KLIP-HuMaCo). Part of the experimental work was conducted with support from the Core Facility Platform of Computer Science and Communication, SIST, ShanghaiTech University.
}


\bibliographystyle{abbrv-doi-hyperref}

\balance
\bibliography{template}

\appendix 

\end{document}


\appendix

\section{Design Alternative}
\par During the iterative design process, we explored three visualization alternatives to balance clarity, scalability, and analytical utility. The first (\cref{fig:G-alter}-(A)) employs a \textit{Circular Dendrogram} where one-hop entities are arranged circumferentially around interpretative path entities, grouped by type for clarity. Hypothesis-aligned entities (lacking direct connections) are placed below. However, increasing one-hop entities caused overlapping edges and visual clutter, particularly between opposing arcs. This obscured relationships and increased cognitive load during analysis. The second design (\cref{fig:G-alter}-(B)) divides the dendrogram into upper and lower arcs, isolating hypothesis-aligned entities. Although this design reduces edge crossings, it retains scalability issues as node counts grow. Uniform node sizing also failed to convey connection density, and expanding to multi-hop entities exacerbated spatial inefficiency. The final approach (\cref{fig:G-alter}-(C)) utilizes the \textit{Voronoi Treemap}, which employs irregularly shaped regions to eliminate node overlap and minimize edge congestion, with two key features: 1) Node Area Encoding: Size reflects connection density, enabling rapid identification of critical entities; 2) Hierarchical Scalability: Supports multi-hop exploration via vertically stacked treemaps, preserving clarity even as complexity increases.

\begin{figure}[h]
\centering
\vspace{-3mm}
\includegraphics[width=\linewidth]{figs/alternative.pdf}
\vspace{-6mm}
\caption{Design alternatives for Hypothesis View.}
\label{fig:G-alter}
\vspace{-3mm}
\end{figure}

\section{Model Performance}
\subsection{Synthetic Lethality}

\par KR4SL trains by ranking all possible candidate genes for a given primary gene using a multi-class cross-entropy loss. Since there’s no constraint on the number of negative samples during training, the model's objective is to assign higher scores to true SL partner genes, without needing to balance positive and negative pairs explicitly. At test time, the model generates a ranked list of candidate partners for each primary gene based on the predicted scores.

\par To assess the ranking quality of the model’s predictions, we use three commonly adopted metrics: Precision@N, Recall@N, and Normalized Discounted Cumulative Gain at N (NDCG@N).

\par Precision@N reflects the fraction of correctly predicted SL partners among the top N ranked candidates. Recall@N, on the other hand, measures the coverage—how many of the actual SL partners are retrieved within the top N predictions. NDCG@N offers a more nuanced evaluation by incorporating the positions of the correct predictions: correct hits appearing earlier in the ranking contribute more to the score.

\par The NDCG@N metric is computed as:

\begin{equation} \text{NDCG@N} = \frac{\text{DCG@N}}{\text{IDCG@N}} \end{equation}

\par In this equation, DCG (Discounted Cumulative Gain) is calculated as:

\begin{equation} \text{DCG@N} = \sum_{i=1}^{N} \frac{s_i}{\log_2(i + 1)} \end{equation}

\par where $s_i$ indicates whether the gene at rank $i$ is a true SL partner. The ideal DCG (IDCG@N) corresponds to the maximum DCG obtainable when all true partners are ranked at the top.

\par Each of these metrics ranges from 0 to 100\%, and final performance is reported as the average over all primary genes in the test set. Higher scores represent more accurate and effective prediction outcomes.

\begin{table}[h]
\vspace{-3mm}
\caption{Model Performance of Synthetic Lethality Prediction: KR4SL.}
\centering
\vspace{-3mm}
\begin{tabu}{cccc}
\toprule
    & \textbf{NDCG@50}   & \textbf{Precision@50}  & \textbf{Recall@50}\\
\midrule
KG4SL\cite{wang2021kg4sl} & 0.0153 & 0.0563 & 0.0560\\
SLGNN\cite{zhu2023slgnn} & 0.0316 & 0.0865 & 0.0862\\
NSF4SL\cite{wang2022NSF4SL} & 0.2125 & 0.3460 & 0.3466\\
\textbf{KR4SL}\cite{zhang2023kr4sl} & \textbf{0.3663} & \textbf{0.5313} & \textbf{0.5305}\\
\bottomrule
\end{tabu}
\label{tab:table_model_performance}
\vspace{-3mm}
\end{table}

\subsection{Drug Repurposing}

\textbf{Mean Percentile Rank (MPR)} is the average percentile rank of the 3-hop DrugMechDB-matched BKG-based paths for true-positive drug–disease pairs~\cite{ma2023kgml}:

\begin{equation}
\mathrm{MPR} = \frac{1}{|\mathrm{PR}|} \sum_{pr \in \mathrm{PR}} pr
\end{equation}

where $\mathrm{PR}$ is a list of percentile ranks of DrugMechDB-matched BKG-based paths for true-positive drug–disease pairs.

\vspace{1em}

\textbf{Mean Reciprocal Rank (MRR)} is the average inverse rank of true-positive drug–disease pairs or their 3-hop DrugMechDB-matched BKG-based paths:

\begin{equation}
\mathrm{MRR} = \frac{1}{|\mathrm{R}|} \sum_{r \in \mathrm{R}} \frac{1}{r}
\end{equation}

where $\mathrm{R}$ is a list of ranks of true-positive drug–disease pairs (for the DRP task) or DrugMechDB-matched BKG-based paths (for the MOA prediction task).

\vspace{1em}

\textbf{Hit@K} is the proportion of ranks not larger than $K$ among true-positive drug–disease pairs (``treat'' category) or their 3-hop DrugMechDB-matched BKG-based paths:

\begin{equation}
\mathrm{Hit@K} = \frac{1}{|\mathrm{R}|} \sum_{r \in \mathrm{R}} \mathbb{I}(r \leq K)
\end{equation}

where $\mathrm{R}$ is a list of ranks of true-positive drug–disease pairs (for the DRP task) or DrugMechDB-matched BKG-based paths (for the MOA prediction task), and $\mathbb{I}(\cdot)$ is the indicator function.

\begin{table}[h]
\vspace{-3mm}
\caption{Model Performance of MOA prediction: KGML-xDTD.}
\centering
\vspace{-3mm}
\begin{tabu}{cccc}
\toprule
    & \textbf{MPR}   & \textbf{MRR}  & \textbf{Hit@50}\\
\midrule
MultiHop~\cite{lin2018multi} & 61.400\% & 0.027 & 0.067\\
\textbf{KGML-xDTD} & \textbf{94.696\%} & \textbf{0.109} & \textbf{0.496}\\
\bottomrule
\end{tabu}
\label{tab:table_model_performance_XD}
\vspace{-3mm}
\end{table}

\section{Dimensionality Reduction Results Comparison}

\subsection{Synthetic Lethality Data}

\begin{figure}[H]
\centering
\vspace{-3mm}
\includegraphics[width=\linewidth]{figs/PCA.png}
\vspace{-3mm}
\caption{\textit{PCA} result of SL Data.}
\label{pca_sl}
\end{figure}

\begin{figure}[H]
\centering
\vspace{-3mm}
\includegraphics[width=\linewidth]{figs/tsne.png}
\vspace{-3mm}
\caption{\textit{t-SNE} result of SL Data.}
\label{tsne_sl}
\end{figure}

\subsection{Drug Repurposing Data}

\begin{figure}[H]
\centering
\vspace{-3mm}
\includegraphics[width=\linewidth]{figs/PCA_drug.png}
\vspace{-3mm}
\caption{\textit{PCA} result of Drug Repurposing Data.}
\label{pca_drug}
\end{figure}

\begin{figure}[H]
\centering
\vspace{-3mm}
\includegraphics[width=\linewidth]{figs/tsne_drug.png}
\vspace{-3mm}
\caption{\textit{t-SNE} result of Drug Repurposing Data.}
\label{tsne_drug}
\end{figure}


In this study, we explored various dimensionality reduction approaches with the objective of integrating SL data with drug repurposing data. Among the approaches investigated, \textit{PCA}~\cite{wold1987principal} exhibited the fastest computational speed; however, its clustering performance was not prominent (\cref{pca_sl},\cref{pca_drug}). \textit{t-SNE}~\cite{Maaten2008VisualizingDU} required significantly more computational time and showed some improvement in clustering quality; nevertheless, it remained suboptimal for high-dimensional data such as gene profiles (\cref{tsne_sl}), and its ability to distinguish between drugs with similar descriptions was limited (\cref{tsne_drug}). Conversely, \textit{UMAP}~\cite{McInnes2018UMAPUM} demonstrated superior efficacy in reducing dimensionality and facilitating the clustering of both gene (\cref{umap}) and drug data (\cref{umap_drug}) with flexible calibration of parameters. It is noteworthy that UMAP has gained widespread adoption in biomedical research for data visualization purposes~\cite{dorrity2020dimensionality,yang2021dimensionality}. While \textit{UMAP} is a non-linear method—meaning distances in the \textit{Embedding View} are not linearly scaled-it prioritizes the preservation of local relationships over global ones. This emphasis on local structure enhances the clustering of nearby points, making it well-suited for our goal of analyzing the correspondence within local clusters.

\begin{figure}[H]
\centering
\vspace{-3mm}
\includegraphics[width=\linewidth]{figs/umap.png}
\vspace{-3mm}
\caption{\textit{UMAP} result of SL Data.}
\label{umap}
\end{figure}

\begin{figure}[H]
\centering
\vspace{-3mm}
\includegraphics[width=\linewidth]{figs/umap_drug.png}
\vspace{-3mm}
\caption{\textit{UMAP} result of .Drug Repurposing Data.}
\label{umap_drug}
\end{figure}





\section{Prompts used for HypoChainer}
\subsection{STAGE1: Contextual Exploration}

\begin{tcolorbox}[
  title=RAG - Recommend entities,
  colback=myBackColor,
  colframe=myFrameColor,
  coltext=black!80, 
  sharp corners,
  boxrule=0.8pt,
  breakable,
  enhanced,
    before skip=5pt,
  after skip=10pt,
]
\begin{myverbatim}
---Role---
You are a helpful assistant specializing in biomedical literature and structured knowledge graphs. You answer user queries based on the provided Data Sources below.

---Goal---
Given the user’s query and the provided Data Sources, generate a concise and relevant response that identifies key biomedical entities and explains their relevance. Use both structured knowledge (KG) and unstructured documents (DC) to support your answer. Incorporate general biomedical knowledge only when it clearly supports the Data Sources.

The response should be fact-based, context-aware, and aligned with the user’s intent — such as identifying related diseases, mechanisms, drug targets, or biomarkers.

---Response Rules---
- Return your answer in JSON format only.
- For each recommended entity, include:
1.  `entity_name`: the name of the disease or condition (as-is from source)
2. `category`: the type or group the entity belongs to
3. `reason`: why this condition is a good candidate
- Unless the user specifies otherwise, return 5–7 of the most relevant entities. Avoid generalizations or unrelated entities.
- Make the output concise, medically meaningful.

---Conversation History---
{history}

---Data Sources---
1. From Knowledge Graph(KG):
{kg_context}
2. From Document Chunks(DC):
{vector_context}

---Output Format---
{recommend_entity_json_format}
\end{myverbatim}
\end{tcolorbox}

\subsection{STAGE2: Hypothesis Construction}
\begin{tcolorbox}[
  title=LLM - Analyse the selected path,
  colback=myBackColor,
  colframe=myFrameColor,
  coltext=black!80, 
  sharp corners,
  boxrule=0.8pt,
  breakable,
  enhanced,
    before skip=5pt,
  after skip=10pt,
]
\begin{myverbatim}
---Role---
You are a biomedical research assistant capable of interpreting multi-hop knowledge graph paths involving drugs, genes, and diseases. Your task is to generate natural language explanations of possible biological mechanisms or therapeutic hypotheses implied by the provided path, based on both the entities and the specific relationships between them.

---Goal---
Given a multi-hop path extracted from a biomedical knowledge graph (each hop including subject, relationship, and object), Interpret the biological significance and possible mechanistic insights reflected by this path, based on the semantics of the entities and their relationships. 

If applicable, end with a symbolic **Possible Hypothesis Chain**, summarizing the cascade from upstream intervention to downstream phenotype.

---Response Rules---
- For each hop, explain how the relationship connects the subject and object biologically or pharmacologically.
- Where relevant, incorporate basic biomedical knowledge.
- Avoid reasons not logically inferred or not widely known.
- End with a **Possible 3-hop Hypothesis Chain**, considering both entities and relation, formatted like: {hypo_chain_format}
- Return your answer in JSON format only.

---Conversation History---
{history}

---Input---
{selected_interpretable_path}

---Output Format---
{path_output_format}
\end{myverbatim}
\end{tcolorbox}

\begin{tcolorbox}[
  title=RAG - Retrieve entities based on hypothesis, colback=myBackColor, colframe=myFrameColor, coltext=black!80, sharp corners, boxrule=0.8pt, breakable, enhanced, before skip=5pt, after skip=10pt,]
\begin{myverbatim}
---Role---
You are a biomedical assistant specializing in retrieving relevant biomedical entities from a Knowledge Graph (KG) based on user input queries. You can understand natural language hypotheses and identify matching entities using semantic and ontological information.

---Goal---
Given a natural language input that describes a biomedical concept, function, or hypothesis (e.g., drug purpose, disease mechanism, symptom category), search **the structured Knowledge Graph(KG)** and return the most relevant entities.

For each retrieved entity:
- Include its name and category (from KG or inferred).
- Generate a brief description summarizing its biological relevance or defining features.
- The description can incorporate general biomedical knowledge, but should be consistent with the context in the Knowledge Graph.

---Response Rules---
- By default, return **15–20 entities**, unless the user specifies otherwise.
- For each entity, provide:
    - `entity_name`: exact name as in KG.
    - `category`: category provided in KG (e.g. Disease, Drug, Gene)
    - `description`: 1–2 sentence summary of the entity and why it’s relevant to the query.
- Return your answer in JSON format only.
- **Do not** provide any entity not in KG.

---Conversation History---
{history}

---Data Sources---
From Knowledge Graph(KG):
{kg_context}

---Output Format---
{retrieval_entity_json_format}
\end{myverbatim}
\end{tcolorbox}

\begin{tcolorbox}[
  title=LLM - Analyze and improve the 3-hop hypothesis chain,
  colback=myBackColor,
  colframe=myFrameColor,
  coltext=black!80, 
  sharp corners,
  boxrule=0.8pt,
  breakable,
  enhanced,
    before skip=5pt,
  after skip=10pt,
]
\begin{myverbatim}
---Role---
You are a biomedical assistant that helps users reason over multi-hop knowledge graph paths. You not only interpret the biological implications of the user-constructed hypothesis chain, but also assess its logical consistency and suggest improvements.

---Goal---
Given a user-constructed biological path (e.g. a drug → relationship1 → gene → relationship2 → process → relationship3 → disease), your task is to:

1. Analyze the **biological validity and logical consistency** of the chain, based on domain knowledge;
2. Identify **potential issues** in entity choice, relation use;
3. Suggest **specific improvements** (e.g., alternative relations or entities, clarification);
4. If the path is valid or partially valid, generate a concise explanation of the mechanism involved.

You may incorporate well-established biological knowledge but must remain grounded in the given chain and its domain context.

---Response Rules---
Your answer should follow this structure:
1. **Chain Assessment**
    - Is the path biologically logical and consistent?
    - Are the relationships appropriate for connecting these entities?
2. **Biological Interpretation**
    - Offer a brief, mechanism-oriented explanation that captures the biological relevance of the refined or validated chain.
3. **Suggested Improvements**
    - If any entity or relation seems inappropriate or too vague, propose alternatives.

---Conversation History---
{history}

---Output Format---
{analyze_and_improve_output_format}
\end{myverbatim}
\end{tcolorbox}

\subsection{General: Response from general LLM}
\begin{tcolorbox}[
  title=LLM - General response grounded in biological principles,
  colback=myBackColor,
  colframe=myFrameColor,
  coltext=black!80, 
  sharp corners,
  boxrule=0.8pt,
  breakable,
  enhanced,
    before skip=5pt,
  after skip=10pt,
]
\begin{myverbatim}
---Role---
You are a biomedical assistant designed to answer expert-level biological or pharmacological questions based on the user’s question and provided contextual knowledge.

---Goal---
Given the user’s query and relevant background information from prior conversation history and supporting documents or knowledge graph paths, generate an informative, accurate, and concise explanation. Your answer should align with the user’s intent (e.g., understanding a mechanism, evaluating a drug-disease link, or tracing a causal relationship).

---Response Rules---
- Answer should be **grounded** in the provided context, but can incorporate general biomedical knowledge if clearly relevant.
- If the user query is broad or vague, **interpret intent from the conversation history** and respond accordingly.
- The tone should be professional and clear, suitable for biomedical researchers or practitioners.
- No hallucinated facts — explanations must be supported by context or well-known knowledge.

---Conversation History---
{history}

\end{myverbatim}
\end{tcolorbox}

\section{Semi-Structured Interview Design}
\par Each interview session~(\cref{tab:interview}) was conducted in a multi-interviewer format, involving 2–3 members of the research team interviewing a single expert to ensure diverse questioning and in-depth exploration. Prior to each interview, participants signed an informed consent form addressing issues of privacy, ethics, and academic use of the data. The semi-structured interview protocol focused on participants’ experiences, challenges, and expectations regarding AI-based scientific discovery systems. All sessions were audio-recorded to ensure accurate and complete documentation.

\par Interviewers took turns posing questions~(\cref{tab:interview_questions}) and follow-ups, encouraging experts to elaborate on their perspectives regarding the use of KGs, LLMs, GNNs, and other AI technologies in scientific discovery. 

\par Interview recordings were transcribed and thoroughly reviewed to correct transcription errors. We employed thematic analysis to systematically code the interview data. The process began with all researchers reading through the transcripts to establish a shared understanding. Then, two team members segmented the data into meaningful units and assigned initial codes. A second group of researchers reviewed and refined the codes, uncovering deeper patterns and themes. Finally, all members discussed and consolidated the coding results to ensure analytical consistency and reliability.

\par Based on the coded data, we synthesized the key challenges and corresponding design requirements expressed by experts in the context of drug repurposing systems. This analysis laid a solid foundation for addressing our subsequent research questions.

\begin{table*}[h]
  \renewcommand{\arraystretch}{1.4}
  \centering
  \caption{Interview Procedure}
  \label{tab:interview}
  \begin{tabularx}{\textwidth}{p{3.5cm} p{3.5cm} X}
    \toprule
    \textbf{Phase} & \textbf{Duration} & \textbf{Content} \\
    \midrule
    Introduction & 5 min & Introduce the study purpose and obtain informed consent. \\
    Background & 5 min & Understand the expert’s research background and tool usage experience. \\
    Thematic Inquiry & 30--40 min & Explore core themes through guided open-ended questions. \\
    Open Feedback & 5--10 min & Encourage additional comments or unmet needs. \\
    Conclusion & 2 min & Express appreciation and conclude the interview. \\
    \bottomrule
  \end{tabularx}
\end{table*}

\begin{table*}[t]
  \caption{Semi-structured Interview Questions for Exploring Experts’ Challenges and Needs}
  \label{tab:interview_questions}
  \centering
  \begin{tabular}{p{4cm}lp{8cm}}
    \toprule
    \textbf{Aspect} & \textbf{Topic} & \textbf{Question} \\
    \midrule
    Trust and Comprehension of Interpretative Paths (\textbf{C1}) & Usage Experience & Have you used GNNs or KG-based systems to uncover biological mechanisms or interpretative paths? What was your experience? \\
    & Sufficiency of Information & When a system presents an interpretative path, do you feel the information is sufficient to help you understand the relationships between entities? \\
    & Relation Label Ambiguity & Have you encountered situations where entities are linked by the same relation label but differ significantly in meaning? Could you provide an example? \\
    & Important Information & In your opinion, what kind of information is most important in a path explanation? What details do you expect to see? \\
    \midrule
    Challenges in Acquiring Knowledge for Hypothesis Construction (\textbf{C2}) & Knowledge Gathering & When constructing novel hypotheses, how do you typically gather background knowledge or inspiration? Is the process time-consuming? \\
    & Tool Helpfulness & Do you find current literature search tools or KG systems helpful for efficiently finding relevant information? \\
    & Presentation Preferences & If a system could recommend mechanisms, documents, or relevant insights, how should it present information to support your hypothesis construction? \\
    \midrule
    Burden of Handling Large-Scale Predictions (\textbf{C3}) & Prediction Usage & Have you tried using model predictions to identify new research focus? How did you process the results? \\
    & Filtering Strategies & When faced with hundreds or thousands of candidate results, what strategies do you use to filter them? What are the challenges? \\
    & Desired Features & If a system could help prioritize or visualize predictions, what kind of features (e.g., ranking, clustering, visual summaries) would be most helpful? \\
    \midrule
    Difficulties in Building Hypothesis Chains (\textbf{C4}) & Reasoning Approach & When constructing a hypothesis chain, which is a logical sequence connecting multiple steps or mechanisms, how do you approach reasoning and verification? \\
    & Expansion Challenges & What do you find most difficult when trying to expand or refine a hypothesis chain? \\
    & System Assistance & If a system could assist with step-by-step hypothesis construction, validation, and refinement, what kind of suggestions or feedback would you expect? \\
    \midrule
    Misalignment Between Predictions and Hypotheses (\textbf{C5}) & Retrieval Methods & Once you have formulated a hypothesis, how do you retrieve relevant predictions or evidence to support it? \\
    & Search Rule Effectiveness & Have you encountered mismatches between your hypothesis and available predictions? Have you tried building custom search rules? How effective were they? \\
    & Automated Retrieval & If the system could automatically retrieve predictions aligned with your hypothesis chain, how would you like that process to work? What level of control or explanation would you need? \\
    \midrule
    Information Gaps Between LLMs, KGs, and Experts (\textbf{C6}) & Integration Experience & Have you tried combining LLMs and KGs to support reasoning or knowledge acquisition? How was your experience? \\
    & Utilization Limitations & When using LLMs, do you feel they fully utilize KG information? If not, in what ways do they fall short? \\
    & Future Expectations & What are your expectations for how future systems could improve the flow and fidelity of information between LLMs, KGs, and human experts? \\
    \bottomrule
  \end{tabular}
\end{table*}

\newpage

\section{Baseline System}
\par The \textit{Baseline} system~(\cref{Baseline}) replaced all RAG components with a standard LLM to create an LLM-only architecture for ablation analysis, enabling comparison of the impact of RAG and KG knowledge within the overall workflow. The Hypothesis View was also removed, with its interface space reallocated to the \textit{Chatbot} to reduce system complexity, allowing us to examine how the absence of this view and the RAG switching mechanism affected both system complexity and overall effectiveness, while keeping the workflow structure intact for task completion.

\begin{figure}[H]
\centering
\vspace{-3mm}
\includegraphics[width=\linewidth]{figs/Group-762.pdf}
\vspace{-3mm}
\caption{\textit{Baseline} System in the User Study.}
\label{Baseline}
\end{figure}

\newpage

\bibliographystyle{abbrv-doi-hyperref}

\balance
\bibliography{template}